\documentclass[11pt,english]{llncs}

\usepackage{amsfonts}

\newcommand{\gilt}{:}
\newcommand{\sodass}{\,:\,}
\newcommand{\setGilt}[2]{\left\{ #1\sodass #2\right\}}

\newcommand{\realrange}[2]{\left[#1, #2\right]}

\newcommand{\unitrange}[2]{\realrange{0}{1}}

\newcommand{\llabel}[1]{\label{\labelprefix:#1}}
\newcommand{\labelprefix}{} 

\newcommand{\discussionsize}{\small}

\newcommand{\frage}[1]{}

\newenvironment{code}{\noindent
\begin{tabbing}%
\hspace{2em}\=\hspace{2em}\=\hspace{2em}\=\hspace{2em}\=\hspace{2em}\=%
\hspace{2em}\=\hspace{2em}\=\hspace{2em}\=\hspace{2em}\=\hspace{2em}\=%
\kill}{\end{tabbing}}

\newcommand{\labelcommand}{}
\newcommand{\captiontext}{}
\newsavebox{\codeparam}
\newcounter{lineNumber}
\newenvironment{disscodepos}[3]{%
\renewcommand{\labelcommand}{#2}%
\renewcommand{\captiontext}{#3}%
\sbox{\codeparam}{\parbox{\textwidth}{#3}}%
\begin{figure}[#1]\begin{center}\begin{code}\setcounter{lineNumber}{1}}{%
\end{code}\end{center}\caption{\llabel{\labelcommand}\captiontext}\end{figure}}

{\end{disscodepos}}

\newcommand{\Is}       {:=}

\newdimen\endofsize\endofsize=0.5em
\def\endofbeweis{~\quad\hglue\hsize minus\hsize
                 \hbox{\vrule height \endofsize width
\endofsize}\par}

\usepackage[T1]{fontenc}
\usepackage[utf8]{inputenc}
\usepackage{color}
\definecolor{note_fontcolor}{rgb}{0.800781, 0.800781, 0.800781}
\usepackage{multirow}
\usepackage{amsfonts}
\usepackage{times}
\usepackage{amsmath}
\usepackage{amssymb}
\usepackage{algorithmic}
\usepackage{algorithm}
\usepackage{mathtools}
\usepackage{wrapfig}
\usepackage{fullpage}
\usepackage{babel}
\usepackage{url}
\usepackage{numprint}
\usepackage{tikz}
\usepackage{pgfplots}
\pgfplotsset{compat=1.12}

\def\MdR{\ensuremath{\mathbb{R}}}

\newcommand{\ie}{i.e.\ }
\newcommand{\eg}{e.g.\ }
\newcommand{\etal}{et~al.\ }
\newcommand{\expansion}{\mathrm{expansion}}

\newcommand{\csch}[2][says]{\niceremark{Christian}{#1}{#2}}
\renewcommand{\csch}[2][says]{}
\newif\ifDoubleBlind
\DoubleBlindfalse

\npdecimalsign{.} 

\date{}

\begin{document}
\newcommand{\mytitle}{Evolutionary Acyclic Graph Partitioning}
\title{\mytitle}
\ifDoubleBlind
\else
\author{Orlando Moreira, Merten Popp and Christian Schulz\\
        \ \\
\textit{Intel Corporation, Eindhoven, The Netherlands}, \\\texttt{\{orlando.moreira, merten.popp\}@intel.com} \\
        \ \\
\textit{Karlsruhe Institute of Technology, Karlsruhe, Germany} \\\textit{and University of Vienna, Vienna, Austria} \\
\texttt{christian.schulz@\{kit.edu, univie.ac.at\}}}
\institute{}
\fi{}

\maketitle
\begin{abstract}
\vspace*{-.5cm}
Directed graphs are widely used to model data flow and execution dependencies in streaming applications. This enables the utilization of graph partitioning algorithms for the problem of parallelizing computation for multiprocessor architectures.  However due to resource restrictions, an acyclicity constraint on the partition is necessary when mapping streaming applications to an embedded multiprocessor.  Here, we contribute a multi-level algorithm for the acyclic graph partitioning problem.  Based on this, we engineer an evolutionary algorithm to further reduce communication cost, as well as to improve load balancing and the scheduling makespan on embedded multiprocessor~architectures. 
\end{abstract}
\section{Practical Motivation}
\vspace*{-.25cm}
\setcounter{page}{1}
\label{intro}

Computer vision and imaging applications have high demands for computational power.
However, these applications often need to run on
embedded devices with severely limited compute resources and a tight thermal
budget. This requires the use of specialized hardware and a programming model
that allows to fully utilize the compute resources.

\ifDoubleBlind
\else
The context of this research is the development of specialized processors at
Intel Corporation for advanced imaging and computer vision. 
\fi{}
In particular, our
target platform is a heterogeneous multiprocessor architecture that is
currently used in Intel processors. Several VLIW processors with vector units
are available to exploit the abundance of data parallelism that typically
exists in imaging algorithms. The architecture is designed for low power and
typically has small local program and data memories.
To cope with memory constraints, it is necessary to break the application,
which is given as a directed dataflow graph, into smaller blocks that are
executed one after another. The quality of this partitioning has a strong
impact on performance. It is known that the problem
is NP-complete \cite{prevWork} and that there is no constant factor approximation algorithm
 for general graphs \cite{prevWork}. Therefore
heuristic algorithms are used in~practice.

We contribute
(a) a new multi-level approach for the acyclic graph partitioning problem, 
(b) based on this, a coarse-grained distributed evolutionary algorithm,
(c) an objective function that improves load balancing on the multiprocessor architecture
and (d) an evaluation on a large set of graphs and a~real~application.
Our focus is on solution quality, not algorithm running time, since these
partitions are typically computed once before the application is compiled.
The rest of the paper is organized as follows: we present all necessary background information on the application graph
and hardware in Section~\ref{background} and then briefly introduce
the notation and related work in
Section~\ref{preliminaries}. Our new multi-level approach is described in Section~\ref{multilevelapproach}. We illustrate the evolutionary algorithm that provides multi-level recombination and mutation operations, as well as a novel fitness function in Section~\ref{evolutionary}. The experimental evaluation of
our algorithms is found in Section~\ref{evaluation}. We conclude in
Section~\ref{conclusion}.

\vspace*{-.25cm}
\section{Background}
\vspace*{-.25cm}
\label{background}
Computer vision and imaging applications can often be expressed as stream
graphs where nodes represent tasks that process the stream data and edges
denote the direction of the dataflow. Industry standards like OpenVX
\cite{openvx} specify stream graphs as Directed Acyclic Graphs (DAG).
In this work, we address the
\begin{wrapfigure}{r}{7cm}
\centering
\vspace*{-.75cm}
\includegraphics[width=7cm]{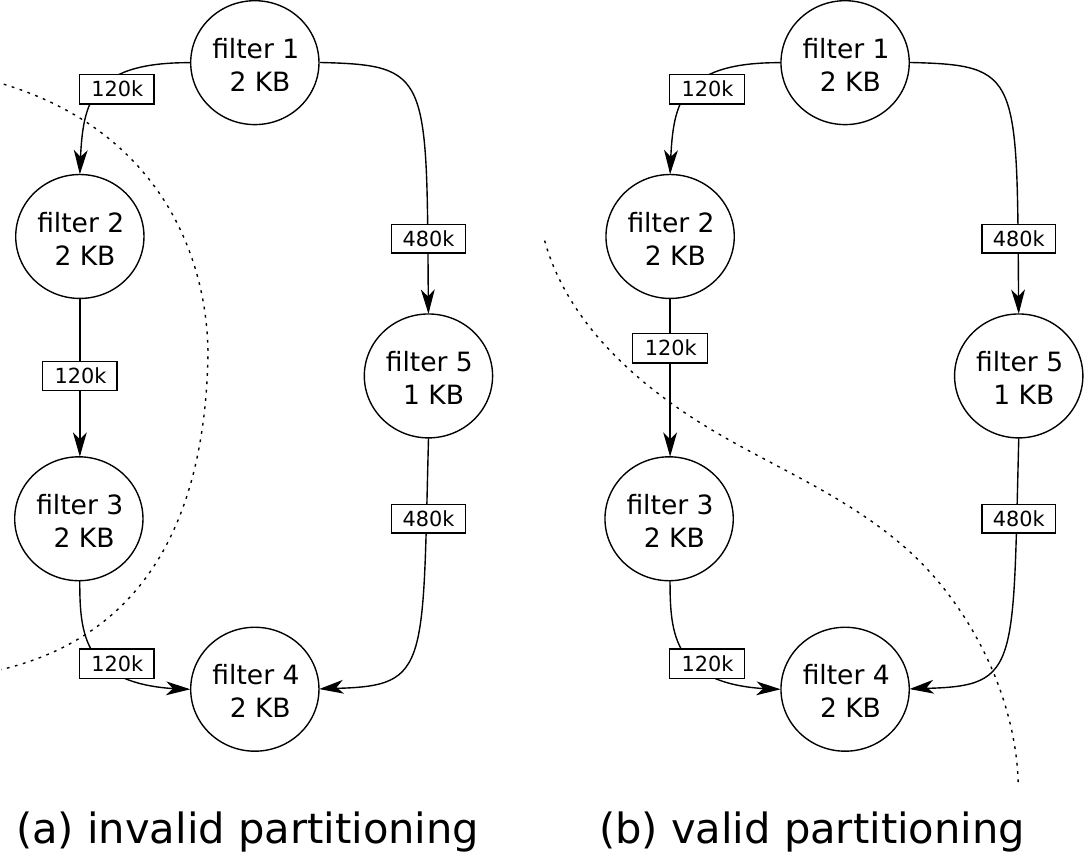}
\caption{Subfigure (a) shows an invalid partition with minimal edge cut, but a
bidirectional connection between blocks and thus a cycle in the quotient graph.
A valid partitioning with minimal edge cut is shown in (b).\label{fig:part_example}}
\vspace*{-.75cm}
\end{wrapfigure}
problem of mapping the nodes of a
\emph{directed acyclic stream graph} to the processors of an embedded multiprocessor.
The nodes of the graph are \emph{kernels} (small, self-contained functions)
that are annotated with code size while edges are annotated with the amount of
data that is transferred during one execution of~the~application.

The processors of the hardware platform have a private local data memory and a
separate program memory. A direct memory access controller (DMA) is used to transfer
data between the local memories and the external DDR memory of the system.
Since the data memories only have a size in the order of hundreds of kilobytes
they can only store a small portion of the image. Therefore the input image is
divided into \emph{tiles}. The mode of operation of this hardware usually is
that the graph nodes are assigned to processors and process the tiles one after
the other.
However, this is only possible if the program memory size is sufficient to
store all kernel programs. For the hardware platform under consideration it
was found that this is not the case for more complex applications such as a
Local Laplacian filter~\cite{paris2011local}. Therefore a gang scheduling
\cite{feitelson1992gang} approach is used where the kernels are divided into
groups of kernels (referred to as gangs) that do not violate memory
constraints. Gangs are executed one after another on the hardware.
After each execution, the kernels of the next gang are loaded. At no time any
two kernels of different gangs are loaded in the program memories of the
processors at the same time. Thus all intermediate data that is produced by
the current gang but is needed by a kernel in a later gang needs to be
transferred to external memory.

A strict ordering of gangs
is required, because data can only be consumed in the same gang where
it was produced and in gangs that are scheduled at a later point in time. If
this does not hold, there is no valid temporal order in which the gangs can be
executed on the platform. Such a partitioning is called \emph{acyclic} because the
quotient graph, which is created by contracting all nodes that are assigned to
the same gang into a single node, does not contain a cycle for a valid acyclic
partitioning.
Figure~\ref{fig:part_example} shows an example for an invalid and a valid
partitioning.

Memory transfers, especially to external memories, are expensive in terms of
power and time. Thus it is crucially important how the assignment of kernels
to gangs is done, since it will affect the amount of data that needs to be
transferred. 
\ifDoubleBlind
\else
In this work, we develop a multi-level approach to enhance our
previous results. 
\fi{}
\vspace*{-.25cm}
\section{Preliminaries}
\vspace*{-.25cm}
\label{preliminaries}
\subsubsection*{Basic Concepts.}
Let $G=(V=\{0,\ldots, n-1\},E,c,\omega)$ be a directed graph
with edge weights $\omega: E \to \MdR_{>0}$, node weights
$c: V \to \MdR_{\geq 0}$, $n = |V|$, and $m = |E|$.
We extend $c$ and $\omega$ to sets, i.e.,
$c(V')\Is \sum_{v\in V'}c(v)$ and $\omega(E')\Is \sum_{e\in E'}\omega(e)$.
We are looking for \emph{blocks} of nodes $V_1$,\ldots,$V_k$
that partition $V$, i.e., $V_1\cup\cdots\cup V_k=V$ and $V_i\cap V_j=\emptyset$
for $i\neq j$.
We call a block $V_i$ \emph{underloaded} [\emph{overloaded}] if
$c(V_i) < L_{\max}$ [if $c(V_i) > L_{\max}$]. 
If a node $v$ has a neighbor in
a block different of its own block then both nodes are called \emph{boundary
nodes}.
An abstract view of the partitioned graph is the so-called \emph{quotient
graph}, in which nodes represent blocks and edges are induced by connectivity
between blocks.
The \emph{weighted} version of the quotient graph has node weights which are
set to the weight of the corresponding block and edge weights which are equal
to the weight of the edges that run between the respective blocks.

A matching $M\subseteq E$ is a set of edges that do not share any common nodes,
i.e., the graph $(V,M)$ has maximum degree one. \emph{Contracting} an edge
$(u,v)$ means to replace the nodes $u$ and $v$ by a new node $x$ connected to
the former neighbors of $u$ and $v$, as well as connecting nodes that have $u$ and
$v$ as neighbors to $x$. 
We set $c(x)=c(u)+c(v)$ so the weight of a node at each level is the number of
nodes it is representing in the original graph. If replacing edges of the form
$(u,w)$,$(v,w)$ would generate two parallel edges $(x,w)$, we insert a single
edge with
$\omega((x,w))=\omega((u,w))+\omega((v,w))$.
\emph{Uncontracting} an edge $e$ undoes its contraction. 
In order to avoid tedious notation, $G$ will denote the current state of the
graph before and after a (un)contraction unless we explicitly want to refer to
different states of the graph.

\vspace*{-.25cm}
\paragraph*{Problem Definition.}
In our context, partitions have to satisfy two constraints: a
balancing constraint and an acyclicity constraint. The \emph{balancing
constraint} demands that
$\forall i\in \{1..k\}\gilt c(V_i) \leq L_{\max} := (1+\epsilon)\lceil\frac{c(V)}{k}\rceil$
for some imbalance parameter $\epsilon \geq 0$.
The \emph{acyclicity constraint} mandates that
the quotient graph is acyclic.
The objective is to minimize the total \emph{edge cut} $\sum_{i,j}w(E_{ij})$
where $E_{ij}\Is\setGilt{(u,v)\in E}{u\in V_i,v\in V_j}$.
The \emph{directed graph partitioning problem with acyclic quotient graph (DGPAQ)}
is then defined as finding a partition $\varPi:=\left\{V_{1,}\ldots,V_{k}\right\}$
that satisfies both constraints while minimizing the objective function. In the \emph{undirected} version of the problem the graph is undirected and no acyclicity constraint is given.

\vspace*{-.25cm}
\paragraph*{Multi-level Approach.}
The multi-level approach to \emph{undirected} graph partitioning consists of three main
phases.
In the \emph{contraction} (coarsening) phase, the algorithm iteratively
identifies matchings $M\subseteq E$ and contracts the edges in $M$. The
result of the contraction is called a \emph{level}. 
Contraction should quickly reduce the size of the input graph and each computed
level should reflect the global structure of the input network. Contraction is
stopped when the graph is small enough to be directly
partitioned.
In the \emph{refinement} phase, the matchings are iteratively
uncontracted. After uncontracting a matching, a refinement algorithm moves
nodes between blocks in order to improve the cut size or balance. 
The intuition behind this approach is that a good partition at one level will
also be a good partition on the next finer level, so local search~converges~quickly. 

\vspace*{-.35cm}
\subsubsection*{Relation to Scheduling.}

Graph partitioning is a sub-step in our scheduling heuristic for the target
hardware platform. We use a first pass of the graph partitioning heuristic with
$L_{\max}$ set to the size of the program memory to find a good composition of
kernels into programs with little interprocessor communication.
The resulting quotient graph is then used in a second pass where $L_{\max}$ is
set to the total number of processors in order to find scheduling gangs that
minimize external memory transfers. In this second step the acyclicity
constraint is crucially important. Note that in the first pass, the constraint
can in principle be dropped. However, this yields programs with interdependencies
that need to be scheduled in the same gang during the second pass. We found
that this often leads to infeasible inputs for~the~second~pass.

While the balancing constraint ensures that the size of the programs in a
scheduling gang does not exceed the program memory size of the platform,
reducing the edge cut will improve the memory bandwidth requirements of the
application.
The memory bandwidth is often the bottleneck, especially in embedded systems. A
schedule that requires a large amount of transfers will neither yield a good
throughput nor good energy efficiency~\cite{panda2001data}.
\ifDoubleBlind
\else
However, in our previous work, we found that our graph partitioning heuristic
while optimizing edge cut occasionally makes a bad decision concerning the
composition of gangs. 
\fi{}
Ideally, the programs in a gang all have equal execution
times. If one program runs considerably longer than the other programs, the
corresponding processors will be idle since the context switch is synchronized.
In this work, we try to alleviate this problem by using a fitness function
in the evolutionary algorithm that considers the estimated execution times of
the~programs~in~a~gang.

After partitioning, a schedule is generated for each gang.
Since partitioning is the focus of this paper, we only give a brief outline.
The scheduling heuristic is a single
appearance list scheduler (SAS). 
In a SAS, the code of a function is never duplicated, in particular, a kernel
will never execute on more than one processor. The reason for using a SAS is
the scarce program memory. List schedulers iterate over a fixed priority list
of programs and start the execution if the required input data and hardware
resources for a program are available. We use a priority list sorted by the
maximum length of the critical path which was calculated
with~estimated~execution~times. Since kernels perform mostly
data-independent calculations, the execution time can be accurately predicted
from the input size which is known from the stream graph and schedule.

\vspace*{-.25cm}
\subsubsection*{Related Work.}
\label{relatedwork}
There has been a vast amount of research on the \emph{undirected} graph partitioning problem so that we
refer the reader to \cite{schloegel2000gph,GPOverviewBook,SPPGPOverviewPaper}
for most of the material.
Here, we focus on issues closely related to our main contributions.
All general-purpose methods for the undirected graph partitioning problem that are able to obtain good partitions for large
real-world graphs are based on the multi-level principle. The basic idea can be
traced back to multigrid solvers for systems of linear equations
\cite{Sou35} but more recent practical methods are based on mostly graph
theoretical aspects, in particular edge contraction and local search. 
For the \emph{undirected} graph partitioning problem, there are many ways to
create graph hierarchies such as matching-based schemes
\cite{Walshaw07,karypis1998fast,Scotch} or variations thereof~\cite{Karypis06}
and techniques similar to algebraic multigrid, \eg{}\cite{meyerhenke2006accelerating}.
However, as node contraction in a DAG can introduce cycles, these methods can \emph{not}
be directly applied to the DAG partitioning problem.
Well-known software packages for the undirected graph partitioning problem that
are based on this approach include Jostle~\cite{Walshaw07},
KaHIP~\cite{kaffpa}, Metis~\cite{karypis1998fast} and Scotch~\cite{ptscotch}.
However, none of these tools can partition directed graphs under the constraint
that the quotient graph is a DAG\@.
Very recently, Hermann \etal\cite{herrmann2017acyclic} presented the first multi-level partitioner for DAGs.
The algorithm finds matchings such that the contracted graph remains acyclic and uses an algorithm comparable to Fiduccia-Mattheyses algorithm \cite{fiduccia1982lth} for refinement.
Neither the code nor detailed results per instance are available at the moment.

Gang scheduling was originally introduced to efficiently schedule parallel
programs with fine-grained interactions~\cite{feitelson1992gang}.
In recent work, this concept has been applied to schedule parallel applications
on virtual machines in cloud computing~\cite{stavrinides2016scheduling}
and extended to include hard real-time tasks~\cite{goossens2016optimal}.
An important difference to our work is that in gang scheduling all tasks that
exchange data with each other are assigned to the same gang, thus there is no
communication between gangs. In our work, the limited program memory of embedded
platforms does not allow to assign all kernels to the same gang.
Therefore, there is communication between gangs which we aim to minimize by
employing graph partitioning methods.

Another application area for graph partitioning algorithms that does have a
constraint on cyclicity is the \emph{temporal partitioning} in the context of
reconfigurable hardware like field-programmable gate arrays (FPGAs).
These are processors with programmable logic blocks that can be reprogramed and
rewired by the user. In the case where the user wants to realize a circuit
design that exceeds the physical capacities of the FPGA, the circuit netlist
needs to be partitioned into partial configurations that will be realized and
executed one after another. 
The first algorithms for temporal partitioning worked on circuit netlists
expressed as hypergraphs. Now, algorithms usually work on a behavioral level
expressed as a regular directed graph.
The proposed algorithms include list scheduling heuristics
\cite{cardoso2000enhanced} or are based on graph-theoretic theorems like
\emph{max-flow min-cut} \cite{jiang2007temporal}, with objective functions
ranging from minimizing the communication cost incurred by the partitioning
\cite{cardoso2000enhanced,jiang2007temporal} to reducing the length of
the critical path in a partition
\cite{cardoso2000enhanced,kao2015performance}. 
Due to the different nature of the problem and different objectives,
a direct comparison with these approaches is not possible.

The algorithm proposed in \cite{chen2012buffer} partitions a directed, acyclic
dataflow graph under acyclicity constraints while minimizing buffer sizes. The authors
propose an optimal algorithm with exponential complexity that becomes
infeasible for larger graphs and a heuristic which iterates over perturbations
of a topological order. The latter is comparable to our initial partitioning and our
first refinement algorithm. We see in the evaluation that moving to a
multi-level and evolutionary algorithm clearly outperforms this approach.
Note that minimizing buffer sizes is not part of our objective.

\vspace*{-.25cm}
\section{Multi-level Approach to Acyclic Graph Partitioning}
\vspace*{-.125cm}
\label{multilevelapproach}
\begin{wrapfigure}{r}{7cm}
\centering
\vspace*{-.75cm}
\includegraphics[width=7cm]{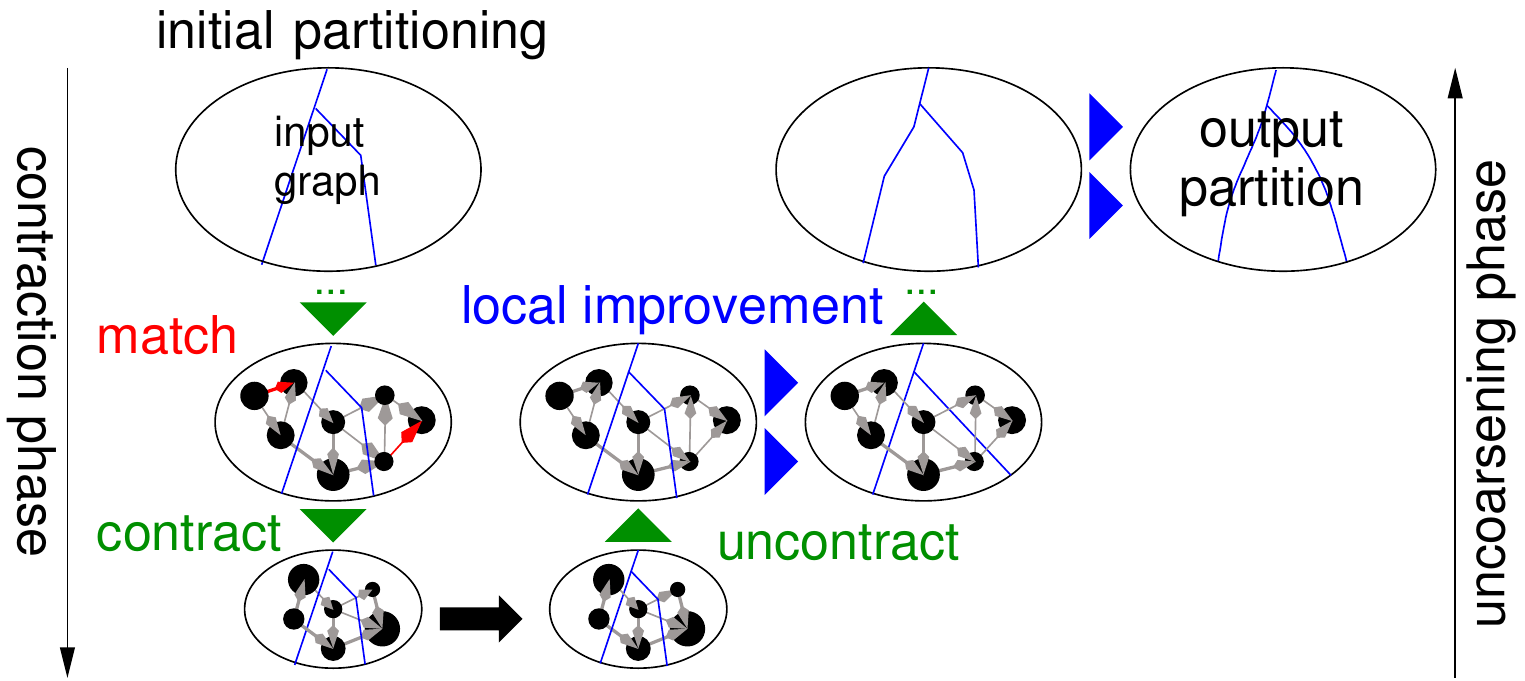}
\caption{The multi-level approach to graph partitioning.}
\vspace*{-.75cm}
\label{fig:multilevelgraphpartitioningapproach}
\end{wrapfigure}
Multi-level techniques have been widely used in the field of graph partitioning
for undirected graphs. We now transfer the techniques used in the
KaFFPa multi-level algorithm~\cite{kaffpa} to a new algorithm that is able to
tackle the DAG partitioning problem.  More precisely, to obtain a multi-level
DAG partitioning algorithm, we integrate local search algorithms that keep the
quotient graph acyclic and handle problems that occur when coarsening a DAG\@. 

Before we give an in-depth description, we present an
overview~of~the~algorithm (see also Figure~\ref{fig:multilevelgraphpartitioningapproach}).
Recall that a multi-level graph partitioner has three phases:
coarsening, initial partitioning and uncoarsening. In \emph{contrast} to
classic multi-level algorithms, our algorithm starts to construct a solution on
the finest level of the hierarchy and not with the coarsening phase.
This is necessary, since contracting matchings can create coarser graphs that
contain cycles, and hence it may become impossible to find feasible solutions
on the coarsest level of the hierarchy.
After initial partitioning of the graph, we continue to coarsen the graph until it
has no matchable edges left. During coarsening, we transfer the solution from the
finest level through the hierarchy and use it as initial partition on the
coarsest graph. As we will see later, since the partition on the finest level
has been feasible, \ie acyclic and balanced, so will be the partition that we
transferred to the coarsest level. The coarser versions of the input graph may
still contain cycles, but local search maintains feasibility on each level and
hence, after uncoarsening is done, we obtain a feasible solution on the finest
level.
The rest of the section is organized as follows. We begin by reviewing the
construction algorithm that we use, continue with the description of the
coarsening phase and then recap local search algorithms for the DAG
partitioning problem that are now used within the multi-level approach.
\vspace*{-.25cm}
\subsection{Initial Partitioning}
\vspace*{-.125cm}
Recall that our algorithm starts with initial partitioning on the finest level
of the hierarchy. 
\ifDoubleBlind
\else
Our initial partitioning algorithm~\cite{prevWork}, 
\fi{}
creates
an initial solution based on a topological ordering of the input graph and
then applies a local search strategy to improve the objective of the solution
while maintaining both constraints -- balance and acyclicity.

More precisely, the initial partitioning algorithm computes a random topological ordering of
nodes using a modified version of Kahn's algorithm with randomized
tie-breaking. The algorithm maintains a list $S$ with all nodes
that have indegree zero and an initially empty list $T$. It then repeatedly
removes a random node $n$ from list $S$, removes $n$ from the graph,
updates $S$ by potentially adding further nodes with indegree zero and adds
$n$ to the tail of $T$. Using list $T$, we can now derive initial solutions by
dividing the graph into blocks of consecutive nodes w.r.t to the ordering. Due
to the properties of the topological ordering, there is no node in a block
$V_{j}$ that has an outgoing edge ending in a block $V_{i}$ with $i<j$. Hence,
the quotient graph of the solution is cycle-free. In addition, the blocks are
chosen such that the balance constraint is fulfilled. The initial
solution is then improved by applying a local search algorithm. Since the construction
algorithm is randomized, we run the heuristics multiple times using different
random seeds and pick the best~solution~afterwards. We call this algorithm
\emph{single-level algorithm}.

\vspace*{-.25cm}
\subsection{Coarsening}
\vspace*{-.125cm}
Our coarsening algorithm makes contraction more systematic by separating two
issues~\cite{kaffpa}: A \emph{rating function} indicates how much sense it
makes to contract an edge based on \emph{local} information. 
A \emph{matching} algorithm tries to maximize the summed ratings of the
contracted edges by looking at the \emph{global} structure of the graph. 
While the rating function allows a flexible characterization of what a ``good''
contracted graph is, the simple, standard definition of the matching problem
allows to reuse previously developed algorithms for weighted matching. 
Matchings are contracted until the graph is ``small enough''. 
In \cite{kappa}, the rating function
$\expansion^{*2}(u,v)\Is \frac{\omega(u,v)^2}{c(u)c(v)}$ works best among other
edge rating functions, so that we also use this rating function for the DAG
partitioning problem.

As in KaFFPa~\cite{kaffpa}, we employ the \emph{Global Path Algorithm
(GPA)} as a matching algorithm. We apply the matching algorithm on the
undirected graph that is obtained by interpreting each edge in the DAG as
undirected edge without introducing parallel edges. The GPA algorithm was proposed by Maue and Sanders~\cite{MauSan07} as a synthesis of the Greedy algorithm and the Path Growing
Algorithm~\cite{DH03a}.
This algorithm achieves a half-approximation in the worst case, but
empirically, GPA gives considerably better results than Sorted Heavy Edge
Matching and Greedy (for more details see \cite{kappa}). 
The GPA algorithm scans the edges in order of decreasing weight but rather than
immediately building a matching, it first constructs a collection of paths and
even cycles, and for each of those computes optimal solutions. 

Recall that our algorithm starts with a partition on the finest level of the
hierarchy. Hence, we set cut edges not to be eligible for the matching
algorithm. This way edges that run between blocks of the given partition are
not contracted. Thus the given partition can be used as a feasible initial partition of
the coarsest graph. The partition on the coarsest level has the same
balance and cut as the input partition.
Additionally, it is also an acyclic partition of the coarsest graph. 
Performing coarsening by this method ensures non-decreasing partition quality,
if the local search algorithm guarantees no worsening. 
Moreover, this allows us to use standard weighted matching algorithms,
instead of using more restricted matching algorithms that ensure that the
contracted graph is also a DAG\@. We stop contraction when no matchable edge is left. 
\vspace*{-.25cm}
\subsection{Local Search}
\vspace*{-.125cm}
\label{sec:localsearch}
Recall that the refinement phase iteratively uncontracts the matchings
contracted during the first phase. Due to the way 
contraction is defined, a partitioning of the coarse level creates a
partitioning of the finer graph with the same objective and balance,
moreover, it \emph{also} maintains the acyclicity constraint on the quotient graph.
After a matching is uncontracted, local search refinement algorithms move nodes
between block boundaries in order to improve the objective while maintaining
the balancing and acyclicity constraint. We use the local search algorithms of Moreira \etal\cite{prevWork}.
We give an indepth description of the algorithms in Appendix~\ref{apdx:localsearch} and shortly outline~them~here. 
All algorithms identify \emph{movable nodes} which can be moved to other
blocks without violating any of the constraints. 
Based on a topological ordering, the first algorithm uses a sufficient condition which can be evaluated quickly to check the acyclicity constraint.
Since the first heuristic can miss possible moves by solely relying upon a
sufficient condition, the second heuristic~\cite{prevWork} maintains a quotient graph during
all iterations and uses Kahn's algorithm to check whether a move creates a cycle in it.
The third heuristic combines the quick check for acyclicity of the first
heuristic with an adapted Fiduccia-Mattheyses algorithm \cite{fiduccia1982lth}
which gives the heuristic the ability to climb out of a local minimum.
\vspace*{-.25cm}
\section{Evolutionary Components}
\vspace*{-.25cm}
\label{evolutionary}
Evolutionary algorithms start with a population of individuals, in our case partitions of the graph, which are created by our multi-level algorithm using different random seeds.
It then evolves the population into different populations over several rounds using recombination and mutation operations. 
In each round, the evolutionary algorithm uses a two-way tournament selection rule \cite{Miller95geneticalgorithms} based on the fitness of the individuals of the population to select good individuals for recombination or mutation. Here, the fittest out of two distinct random individuals from the~population~is~selected.
We focus on a simple evolutionary scheme and generate one offspring per generation. 
When an offspring is generated, we use an eviction rule to select a member of the population and replace it with the new offspring. 
In general, one has to take both, the fitness of an individual and the distance between individuals in the population, into consideration \cite{baeckEvoAlgPHD96}. 
We evict the solution that is \emph{most similar} to the offspring among those individuals in the population that have a cut worse or equal to the cut of the offspring itself. 
The difference of two individuals is defined as the size of the symmetric difference between their sets of cut edges. 

We now explain our multi-level recombine and mutation operators.
Our recombine operator ensures that the partition quality, \ie the edge cut, of the offspring is \emph{at least as good as the best of both parents}.
For our recombine operator, let $\mathcal{P}_1$ and $\mathcal{P}_2$ be two individuals from the population. 
Both individuals are used as input for our multi-level DAG partitioning algorithm in the following sense. 
Let $\mathcal{E}$ be the set of edges that are cut edges, \ie edges that run between two blocks, in either $\mathcal{P}_1$ \emph{or} $\mathcal{P}_2$. 
All edges in $\mathcal{E}$ are blocked during the coarsening phase, \ie they are \emph{not} contracted during the coarsening phase.
In other words, these edges are not eligible for the matching algorithm used during the coarsening phase and therefore are not part of any matching computed.
As before, the coarsening phase of the multi-level scheme stops when no contractable edge is left. 
As soon as the coarsening phase is stopped, we apply the better out of both input partitions w.r.t to the objective to the coarsest graph and use this as initial partitioning. We use random tie-breaking if both input individuals have the same objective value. 
This is possible since we did not contract any cut edge of $\mathcal{P}$.
Again, due to the way coarsening is defined, this yields a feasible partition for the coarsest graph that fulfills both constraints (acyclicity and balance) if the input individuals~fulfill~those.

Note that due to the specialized coarsening phase and specialized initial partitioning, we obtain a high quality initial solution on a very coarse graph. 
Since our local search algorithms guarantee no worsening of the input partition and use random tie breaking, we can assure nondecreasing partition quality.
Also note that local search algorithms can effectively exchange good parts of the solution on the coarse levels by moving only a few nodes. 
Due to the fact that our multi-level algorithms are randomized, a recombine operation performed twice using the same parents can yield a different offspring. 
Each time we perform a recombine operation, we choose one of the local search
algorithms described in Section~\ref{sec:localsearch} uniformly at random.

\vspace*{-.5cm}
\subsubsection*{Cross Recombine.}
This operator recombines an individual of the population with a partition of the graph that can be from a different problem space, \eg a $k'$-partition of the graph.
While $\mathcal{P}_1$ is chosen using tournament selection as before, we create $\mathcal{P}_2$ in the following way.
We choose $k'$ uniformly at random in $[k/4, 4k]$ and $\epsilon'$ uniformly at random in $[\epsilon, 4\epsilon]$. 
We then create $\mathcal{P}_2$ (a $k'$-partition with a relaxed balance constraint) by using the multi-level approach. 
The intuition behind this is that larger imbalances reduce the cut of a partition and using a $k'$-partition instead of $k$ may help us to discover cuts in the graph that otherwise are hard to discover. Hence, this yields good input partitions for our recombine operation. 

\vspace*{-.5cm}
\subsubsection*{Mutation.}
We define two mutation operators. 
Both mutation operators use a random individual $\mathcal{P}_1$ from the current population.
The first operator starts by creating a $k$-partition $\mathcal{P}_2$ using the multi-level scheme. 
It then performs a recombine operation as described above, but not using the better of both partitions on the coarsest level, but $\mathcal{P}_2$.
The second operator ensures nondecreasing quality. It basically recombines $\mathcal{P}_1$ with itself (by setting $\mathcal{P}_2 = \mathcal{P}_1$). 
In both cases, the resulting offspring is inserted into the population using the eviction strategy described above.

\vspace*{-.5cm}
\subsubsection*{Fitness Function.}
\label{sec:fitnessfunction}
Recall that the execution of programs in a gang is synchronized. Therefore, a
lower bound on the gang execution time is given by the longest execution time
of a program in a gang.
Pairing programs with short execution times with a single long-running program
leads to a bad utilization of processors, since the processors assigned to
the short-running programs are idle until all programs have finished.
To avoid these situations, we use a fitness function that estimates the
critical path length of the entire application by identifying the
longest-running programs per gang and summing their execution times.
This will result in gangs, where long-running programs are paired with other
long-running programs.
More precisely, the input graph is annotated with execution times for each node that were
obtained by profiling the corresponding kernels on our target hardware. The
execution time of a program is calculated by accumulating the execution times
for all firings of its contained kernels.
The quality of a solution to the partitioning problem is then measured by the
fitness function which is a linear combination of the obtained edge cut and the
critical path length.
Note, however, that the recombine and mutation operations still optimize for cuts.

\vspace*{-.35cm}
\subsubsection*{Miscellanea.}
We follow the parallelization approach of \cite{kaffpaE}: 
Each processing element (PE) has its own population and performs the same operations using different random seeds. 
The parallelization / communication protocol is similar to \textit{randomized rumor spreading} \cite{conf/icalp/DoerrF11}. 
We follow the description of \cite{kaffpaE} closely: A communication step is organized in rounds. 
In each round, a PE chooses a communication partner uniformly at random among those who did not yet receive $P$ and sends the current best partition $P$ of the local population. 
Afterwards, a PE checks if there are incoming individuals and if so inserts them into the local population using the eviction strategy described above.
If $P$ is improved, all PEs are again eligible.

\vspace*{-.25cm}
\section{Experimental Evaluation}
\vspace*{-.25cm}
\label{evaluation}
\label{s:experiments}

\paragraph*{System.}

\ifDoubleBlind
\else
We have implemented the algorithms described above within the KaHIP~\cite{kaffpa} framework using C++.
\fi{}
All programs have been compiled using g++ 4.8.0 with full optimizations turned on (-O3 flag) and 32 bit index data types.
We use two machines for our experiments: \emph{machine A} has two Octa-Core Intel Xeon E5-2670 processors running at 2.6\,GHz with 64 GB of local memory. We use this machine in Section~\ref{sec:evoexpdagcutojbective}.
\emph{Machine B} is equipped with two Intel Xeon X5670 Hexa-Core processors
(Westmere) running at a clock speed of 2.93 GHz. The machine has 128 GB main
memory, 12 MB L3-Cache and 6$\times$256 KB L2-Cache. We use this machine in Section~\ref{sec:expplatform}.
Henceforth, a PE is one core.
\paragraph*{Methodology.}
We mostly present two kinds of data: average values and plots that show the evolution of solution quality (\textit{convergence plots}).
In both cases we perform multiple repetitions. The number of repetitions is dependent on the test that we perform.
Average values over multiple instances are obtained as follows: for each instance (graph, $k$), we compute the geometric mean of the average edge cut for each instance. 
We now explain how we compute the convergence plots,
starting with how they are computed for a single instance $I$:
whenever a PE creates a partition, it reports a pair ($t$, cut) where the timestamp $t$ is the current elapsed time on the particular PE and \emph{cut} refers to the cut of the partition that has been created.
When performing multiple repetitions, we report average values ($\overline{t}$, avgcut) instead.
After completion of the algorithm, we have $P$ sequences of pairs ($t$, cut) which we now merge into one sequence.
The merged sequence is sorted by the timestamp $t$. 
The resulting sequence is called $T^I$.
Since we are interested in the evolution of the solution quality, we compute another sequence $T^I_{\text{min}}$.
For each entry (in sorted order) in $T^I$ we insert the entry $(t, \min_{t'\leq t} \text{cut}(t'))$ into $T^I_\text{min}$.
Here $\min_{t'\leq t} \text{cut}(t')$ is the minimum cut that occurred until time $t$.
$N^I_{\text{min}}$ refers to the normalized sequence, i.e. each entry ($t$, cut) in $T^I_\text{min}$ is replaced by ($t_n$, cut) where $t_n = t/t_I$ and $t_I$ is the average time that the multi-level algorithm needs to compute a partition for the instance $I$.
To obtain average values over \textit{multiple instances} we do the following: for each instance we label all entries in $N^I_{\text{min}}$, i.e. ($t_n$, cut) is replaced by ($t_n$, cut, $I$). We then merge all sequences $N^I_\text{min}$ and sort by $t_n$. The resulting sequence is called $S$. 
The final sequence $S_g$ presents \textit{event based} geometric averages values. 
We start by computing the geometric mean cut value $\mathcal{G}$ using the first value of all $N^I_\text{min}$ (over $I$).
To obtain $S_g$, we sweep through $S$: for each entry (in sorted order) $(t_n, c, I)$ in $S$ we update $\mathcal{G}$, i.e. the cut value of $I$ that took part in the computation of $\mathcal{G}$ is replaced by the new value $c$, and insert $(t_n, \mathcal{G})$ into $S_g$. 
Note that $c$ can be only smaller or equal to the old cut value of $I$.

\vspace*{-.125cm}
\paragraph*{Instances.}
We use the algorithms under consideration on a set of instances from the
Polyhedral Benchmark suite (PolyBench) \cite{pouchet2012polybench} which have
been kindly provided by Hermann \etal\cite{herrmann2017acyclic}. In addition,
we use an instance of Moreira~\cite{prevWork}.
Basic properties of the instances can be found in Appendix~Table~\ref{tab:test_instances}.

\vspace*{-.25cm}
\subsection{Evolutionary DAG Partitioning with Cut as Objective}
\begin{figure}[t]
\centering
\includegraphics[width=5cm]{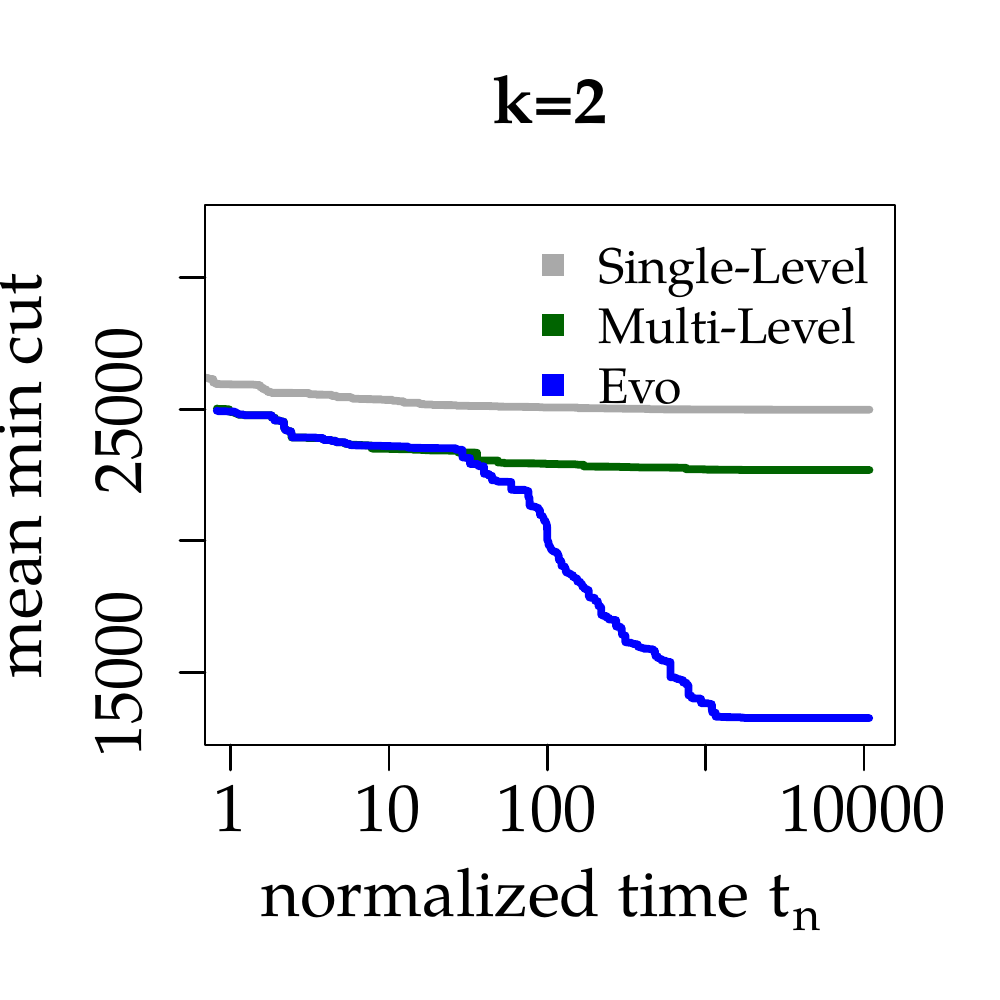} 
\includegraphics[width=5cm]{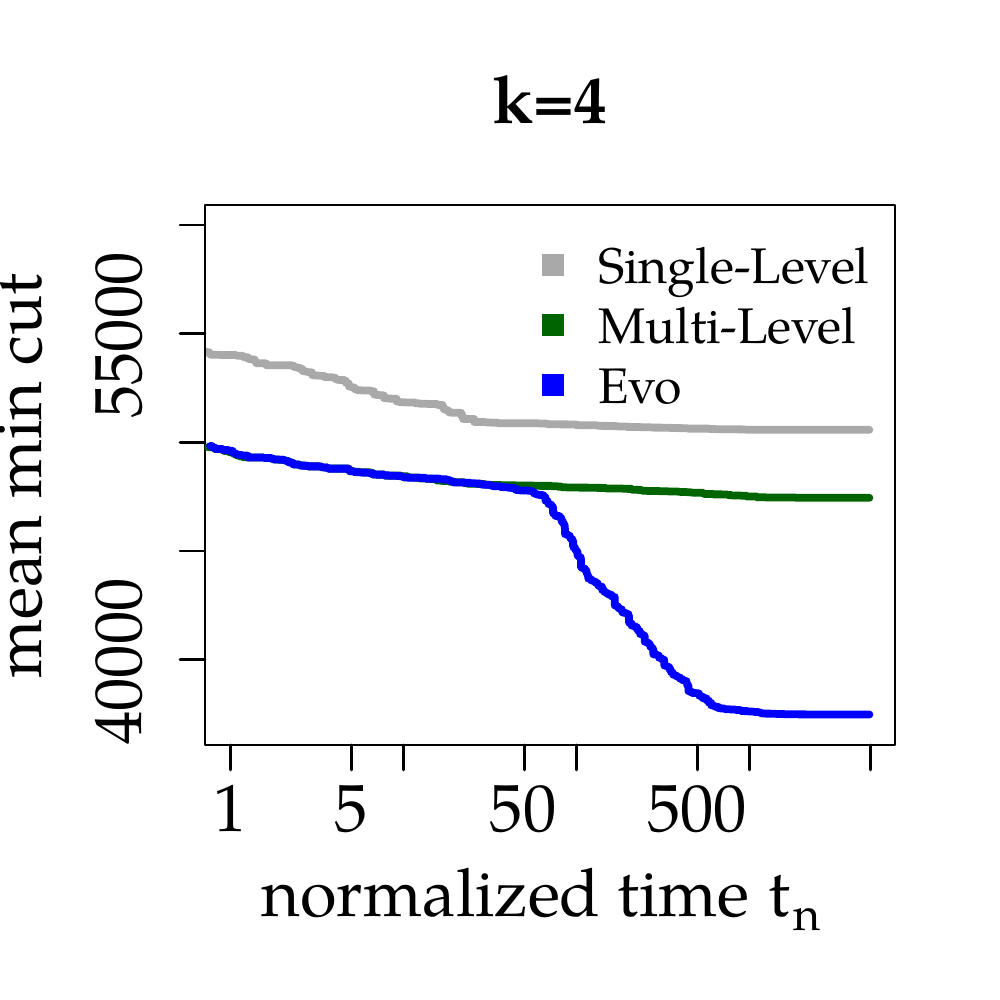} 
\includegraphics[width=5cm]{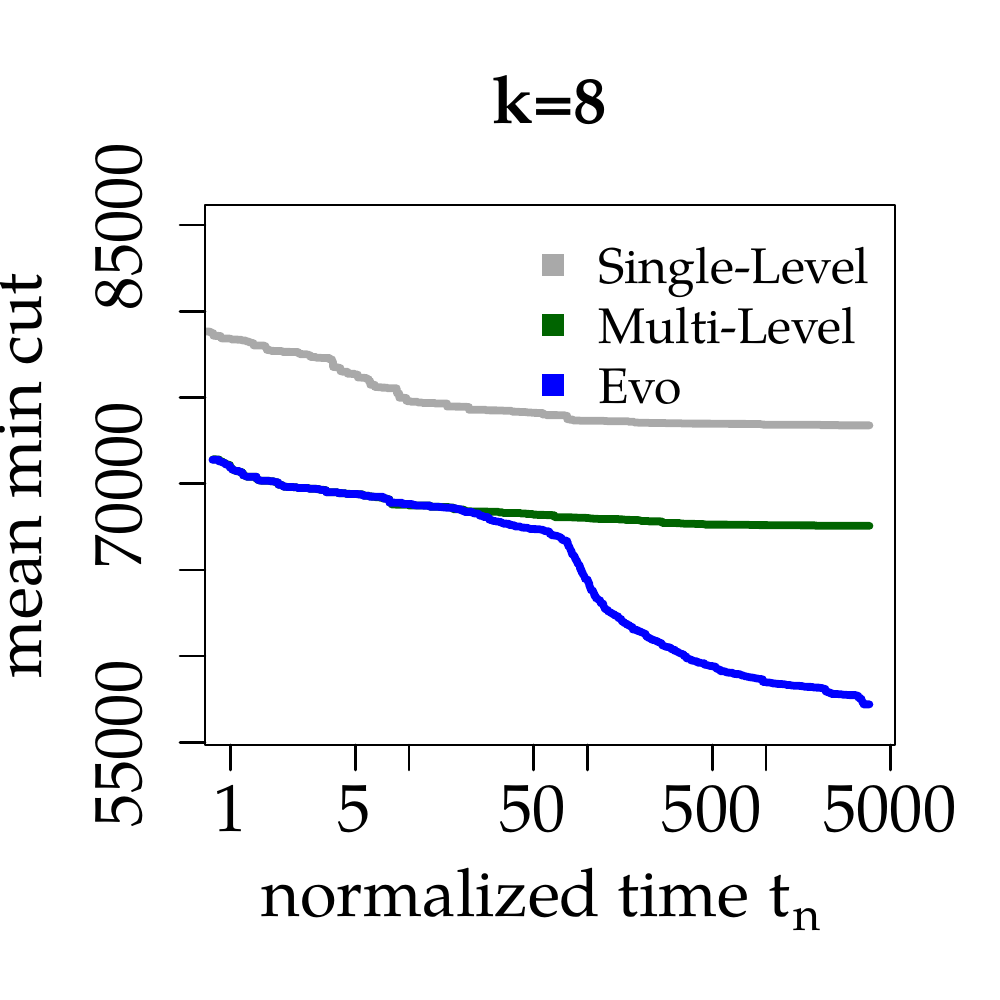}  \\
                       \vspace*{-.25cm}
\includegraphics[width=5cm]{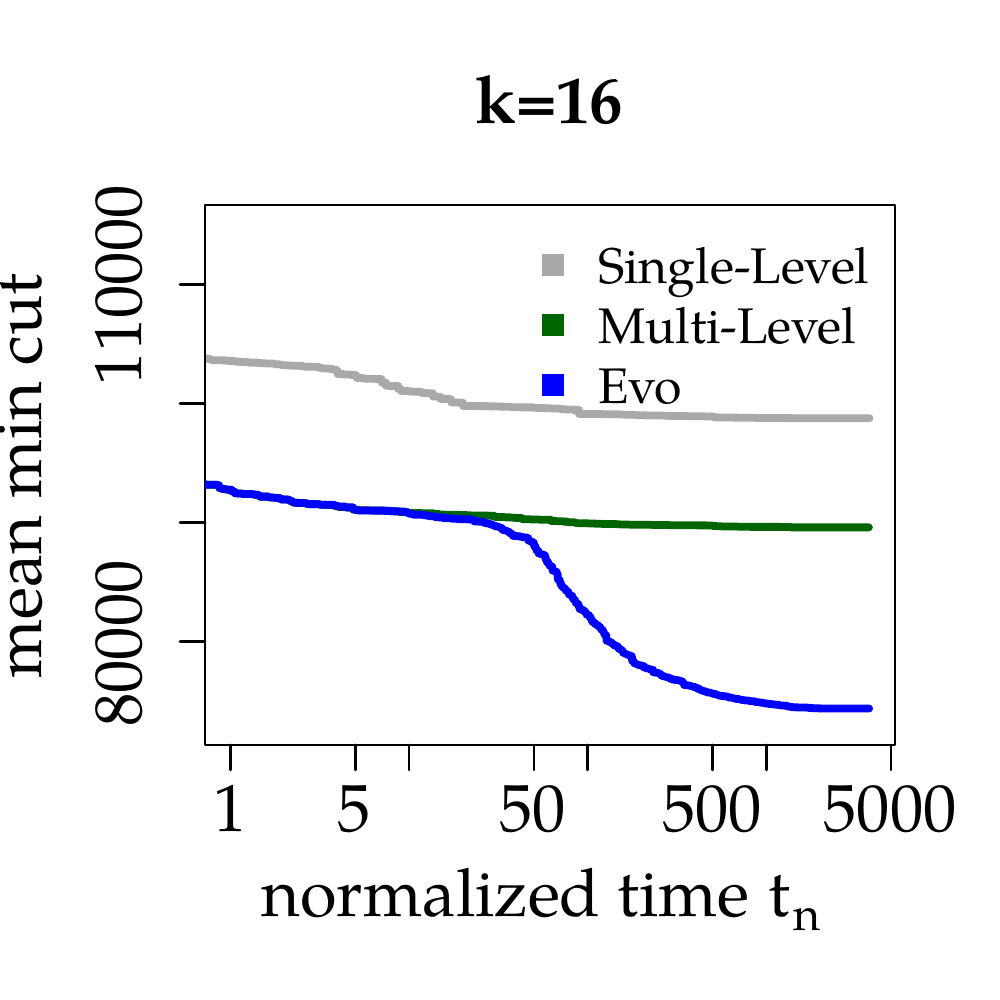} 
\includegraphics[width=5cm]{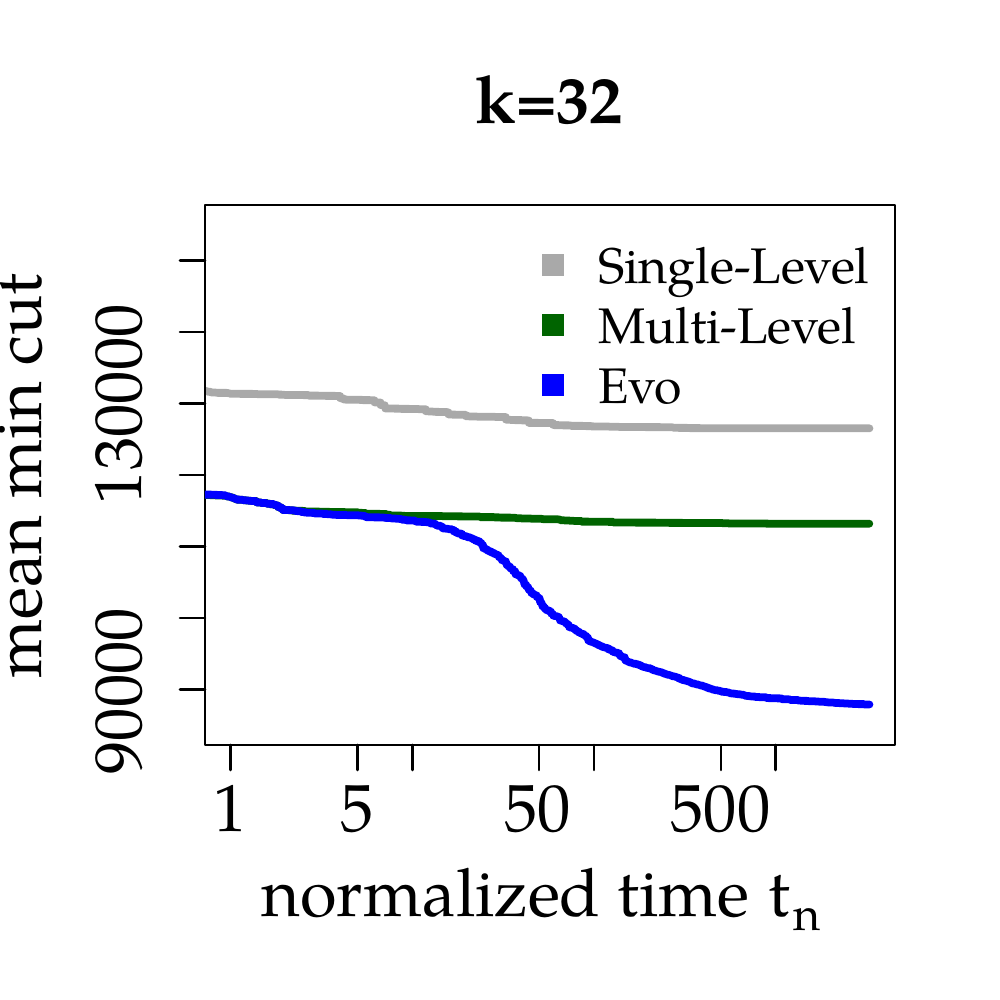} 
\includegraphics[width=5cm]{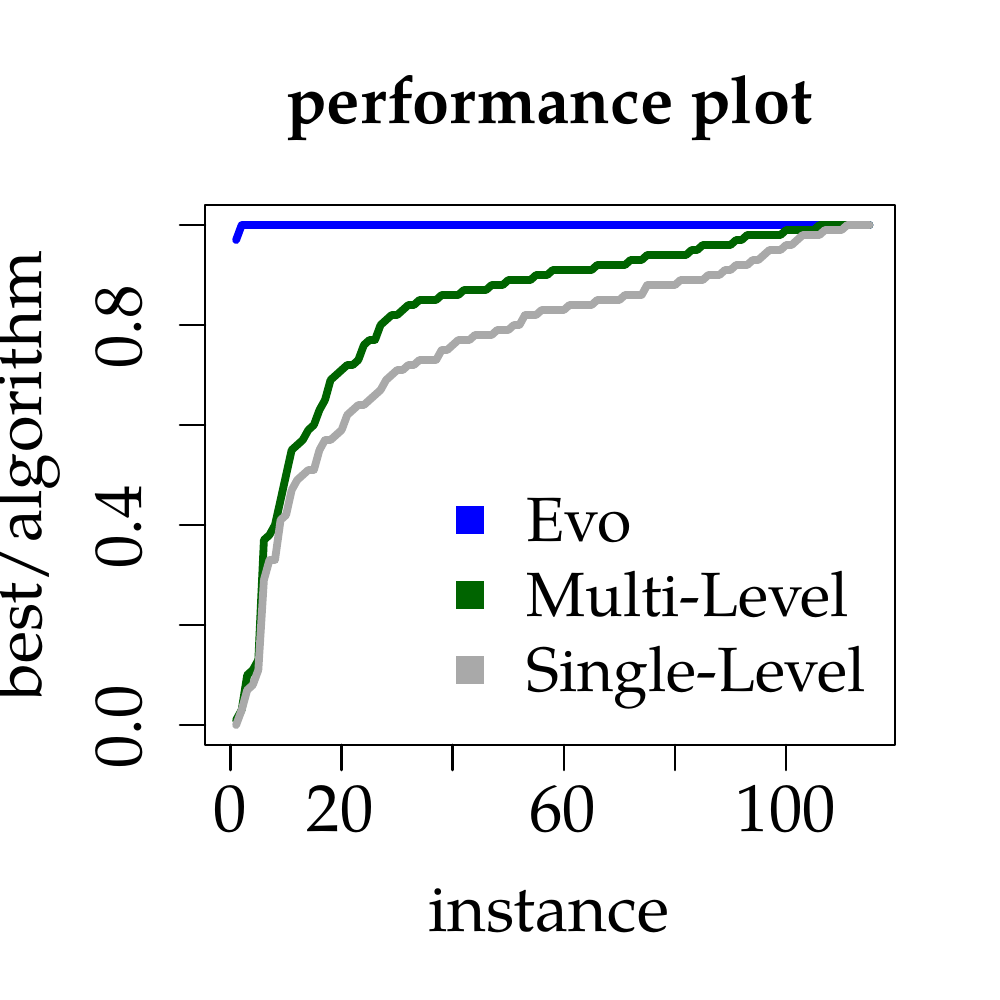}
\vspace*{-.5cm}
\caption{Convergence plots for $k \in \{2,4,8,16,32\}$ and a performance plot.}
\vspace*{-.75cm}
\label{fig:comparision}
\end{figure}

\vspace*{-.125cm}
\label{sec:evoexpdagcutojbective}
We will now compare the different proposed algorithms. 
Our main objective in this section is the cut objective. 
In our experiments, we use the imbalance parameter $\epsilon=3\%$. 
We use 16 PEs of machine A and two hours of time per instance when we use the evolutionary algorithm.
We parallelized repeated executions of multi- and single-level algorithms since they are embarrassingly parallel for different seeds and also gave 16 PEs and two hours of time to each of the algorithms.
Each call of the multi-level and single-level algorithm uses one of our local search algorithms at random and a different random seed.
We look at $k \in \{2,4,8,16,32\}$ and performed three repetitions per instance.
Figure~\ref{fig:comparision} shows convergence and performance plots and Tables~\ref{tab:detailedone}, \ref{tab:detailedtwo} in the Appendix show detailed results per instance.
To get a visual impression of the solution quality of the different algorithms, Figure~\ref{fig:comparision} also presents a \emph{performance plot} using all instances (graph, $k$). 
A curve in a performance plot for algorithm~X is obtained as follows: For each instance, we calculate the ratio between the best cut obtained by any of the considered algorithms and the cut for algorithm~X. These values are then sorted.

First of all, the performance plot in Figure~\ref{fig:comparision} indicates that our evolutionary algorithm finds significantly smaller cuts than the single- and multi-level scheme.
Using the multi-level scheme instead of the single-level scheme already improves the result by 10\% on average. This is expected since using the multi-level scheme introduces a more global view to the optimization problem and the multi-level algorithm starts from a partition created by the single-level algorithm (initialization algorithm + local search).
In addition, the evolutionary algorithm always computes a better result than the single-level algorithm. This is true for the average values of the repeated runs as well as the achieved best cuts. 
The single-level algorithm computes average 
cuts that are 42\% larger than the
 ones computed 
by the evolutionary algorithm and best cuts that are 47\% larger than the best cuts computed by the evolutionary 
algorithm. 
As anticipated, the evolutionary algorithm computes the best result in almost all cases. 
In three cases the best cut is equal to the multi-level, and in three other cases the result of the multi-level algorithm is better (at most 3\%, e.g. for $k=2$, \texttt{covariance}). 
These results are due to the fact that we already use the multi-level algorithm to initialize the population of the evolutionary algorithm. 
In addition, after the initial population is built, the recombine and mutation operations can successfully improve the solutions in the population further and break out of local minima (see Figure~\ref{fig:comparision}).
Average cuts of the evolutionary algorithm 
\begin{wraptable}{r}{6cm}
\centering
\vspace*{-.7cm}
\begin{tabular}{c|cc}
$k$ & Single-Level & Multi-Level \\
\hline
\hline
2  &  112\% &89\%  \\
4  &  34\% & 23\%   \\
8  &  28\% & 18\% \\
16 &  32\% & 20\% \\
32 &  44\% & 28\%  \\
\hline
\hline
\end{tabular}
\caption{Average increase of best cuts over best cuts of evolutionary algorithm.}
\vspace*{-1cm}
\label{tab:perk}
\end{wraptable}
are  29\% smaller than the average cuts computed by the multi-level algorithm (and 33\% in case of best cuts). 
The largest improvement of the evolutionary algorithm over the single- and multi-level algorithm is a factor 39 (for $k=2$, \texttt{3mm0}).
Table~\ref{tab:perk} shows how improvements are distributed over different values of $k$.
Interestingly,
in contrast to evolutionary algorithms for the undirected graph partitioning problem, e.g. \cite{kaffpaE}, improvements to the multi-level algorithm do not increase with increasing $k$. 
Instead, improvements more diversely spread over different values of $k$. 
We believe that the good performance of the evolutionary algorithm is due to a
very fragmented search space that causes local search heuristics to easily get
trapped in local minima, especially since local search algorithms maintain the
feasibility on the acyclicity constraint.
Due to mutation and recombine operations, our evolutionary algorithm escapes those more effectively than the multi- or single-level approach. 

\vspace*{-.25cm}
\subsection{Impact on Imaging Application}
\vspace*{-.125cm}
\label{sec:expplatform}

We evaluate the impact of the improved partitioning heuristic on
an advanced imaging algorithm, the \emph{Local Laplacian filter}.
The Local Laplacian filter is an edge-aware image processing filter.
The algorithm uses the concepts of \emph{Gaussian pyramids} and \emph{Laplacian
pyramids} as well as a point-wise remapping function to enhance image
details without creating artifacts.
A detailed description of the algorithm and theoretical background
is given in~\cite{paris2011local}.
We model the dataflow of the filter as a DAG where nodes are annotated with
the program size and an execution time estimate and edges with the
corresponding data transfer size.
The DAG has $489$ nodes and $631$ edges in total in our configuration.
We use all algorithms (multi-level, evolutionary), the evolutionary with the
fitness function set to the one described in Section~\ref{sec:fitnessfunction}.
\ifDoubleBlind
\else
We compare our algorithms with the best local search heuristic from our previous work~\cite{prevWork}
\fi{}
The time budget given to each heuristic is ten minutes.
The makespans for each resulting schedule are obtained with a cycle-true
compiled simulator of the hardware platform.
We vary the available bandwidth to external memory to assess the impact of edge cut on schedule makespan.
In the following, a bandwidth of $x$ refers to $x$ times the bandwidth available on the real hardware.
The relative improvement in makespan is compared to
\ifDoubleBlind
\else
our previous heuristic in~\cite{prevWork}.
\fi{}

In this experiment, the results in terms of edge cut as well as makespan are similar for the multi-level and the evolutionary algorithm optimizing for cuts, as the filter is fairly small.
However, the new approaches improve the makespan of the application. This is mainly because the reduction of the edge cut
reduces the amount of data that needs to be transferred to external memories.
Improvements range from 1\% to 5\% depending on the available memory bandwidth with high improvements being seen for small memory bandwidths.
For larger memory bandwidths, the improvement in makespan diminishes 
since the pure reduction of communication volume
becomes less important.
Using our new fitness function that incorporates critical path length 
\emph{increases} the makespan by 40\% to 10\% if the memory bandwidth is scarce
(for bandwidths ranging from 1 to 3).
We found that the gangs in this case are almost always memory-limited and thus reducing
communication volume is predominantly important. With more bandwidth available,
including critical path length in the fitness function improves the
makespan by 3\% to 33\% for bandwidths ranging from 4 to 10.
Hence, using the fitness function results in a convenient way to
fine-tune the heuristic for a given memory bandwidth.
For hardware platforms with a scarce bandwidth, reducing the edge cut is the
best. If more bandwidth is available, for example if more than one DMA engine is available,
one can change the factors of the linear combination to gradually reduce the impact of
edge cut in favor of critical path length.

\vspace*{-.25cm}
\section{Conclusion}
\vspace*{-.25cm}
\label{conclusion}

Directed graphs are widely used to model data flow and execution dependencies
in streaming applications which enables the utilization of graph partitioning
algorithms for the problem of parallelizing computation for multiprocessor
architectures.
In this work, we introduced a novel multi-level algorithm as well as the first evolutionary algorithm for the acyclic graph partitioning problem.
Additionally, we formulated an objective function that improves load balancing on the target platform which is then used as fitness function in the evolutionary algorithm.
Extensive experiments over a large set of graphs and a~real~application indicate that the multi-level as well as the evolutionary algorithm significantly improve the state-of-the-art.
Our experiments indicate that the search space has many local minima. Hence, in future work, we want to experiment with relaxed constraints on coarser levels of the hierarchy. Other future directions of research include multi-level algorithms that directly optimize the newly introduced fitness function.
\bibliography{bibliography}

\begin{appendix}
 \pagenumbering{gobble}

 \section{Details on Local Search Algorithms}
 \label{apdx:localsearch}
The first heuristic identifies \emph{movable nodes} which can be moved to other
blocks without violating the constraints.
It uses a sufficient condition to check the acyclicity constraint.
Since the acyclicity constraint was maintained by the previous steps,
a topological ordering of blocks exists such that all edges between
blocks are forward edges w.r.t. to the ordering. Moving a node from one block
to another can potentially turn a forward edge into a back edge. To ensure
acyclicity, it is sufficient to avoid these moves since only then the original
ordering of blocks will remain intact. This condition can be checked very fast
for a node $v\in V_i$. All incoming edges are checked to find the node $u\in
V_A$ where $A$ is maximal.
$A\leq i$ must hold, otherwise the topological ordering already contains a back
edge. If $A < i$, then the node can be moved to blocks preceding $V_i$ up to
and including $V_A$ in the topological ordering without creating a back edge.
This is because all incoming edges of the node will either be internal to block
$V_A$ or are forward edges starting from blocks preceding $V_A$. The same
reasoning can be made for outgoing edges of $v$ to identify block succeeding
$V_i$ that are eligible for a move.
After finding all movable nodes under this condition, the heuristic will choose
the move with the highest gain.
The complexity of this heuristic is $O(m)$ \cite{prevWork}.

Since the first heuristic can miss possible moves by solely relying upon a
sufficient condition, the second heuristic~\cite{prevWork} maintains a quotient graph during
all iterations and uses Kahn's algorithm to check whether a cycle was created
whenever a move causes a new edge to appear in the quotient graph and the
sufficient condition does not give an answer.
The cost is  $O(km)$ if the quotient graph is
sparse.

The third heuristic combines the quick check for acyclicity of the first
heuristic with an adapted Fiduccia-Mattheyses algorithm \cite{fiduccia1982lth}
which gives the heuristic the ability to climb out of a local minimum.
The initial partitioning is improved by exchanging nodes between a pair of
blocks. The algorithm will then calculate the gain for all movable nodes and
insert them into a priority queue. Moves with highest gains are committed if
they do not overload the target block. After each move, it is checked whether
a former internal node in block is now an boundary node, if so, the gain for
this node is calculated and it is inserted into the priority queue. Similarly,
a node that previously was movable might now be locked in its block. In this
case, the node will be marked as locked since searching and deleting the node
in the priority queue has a much higher computational complexity.

The inner pass of the heuristic stops when the priority queue is depleted or
after $2n/k$ moves which did not have a measurable impact on the quality of
obtained partitionings. The solution with best objective that was achieved
during the pass will be returned.
The outer pass of the heuristic will repeat the inner pass for randomly chosen
pairs of blocks. At least one of these blocks has to be ``active''. Initially,
all blocks are marked as ``active''. If and only if the inner pass results in
movement of nodes, the two blocks will be marked as active for the next
iteration. The heuristic stops if there are no more active blocks.
The complexity is $O(m+n\log \frac{n}{k})$ if the quotient graph is~sparse.

 \section{Basic Instance Properties}
\begin{table}[h]
	\centering
        \small
	\begin{tabular}{| l | r | r ||l|r|r| }
			\hline
		 	Graph & $n$ & $m$ & Graph & $n$ & $m$ \\
		 	\hline \hline

2mm0       & \numprint{36500}   & \numprint{62200}  & atax       & \numprint{241730}  & \numprint{385960}\\
syr2k      & \numprint{111000}  & \numprint{180900} &  symm       & \numprint{254020}  & \numprint{440400}\\
3mm0       & \numprint{111900}  & \numprint{214600} &  fdtd-2d    & \numprint{256479}  & \numprint{436580}\\
doitgen    & \numprint{123400}  & \numprint{237000} &  seidel-2d  & \numprint{261520}  & \numprint{490960}\\
durbin     & \numprint{126246}  & \numprint{250993} &  trmm       & \numprint{294570}  & \numprint{571200}\\
jacobi-2d  & \numprint{157808}  & \numprint{282240} &  heat-3d    & \numprint{308480}  & \numprint{491520}\\
gemver     & \numprint{159480}  & \numprint{259440} &  lu         & \numprint{344520}  & \numprint{676240}\\
covariance & \numprint{191600}  & \numprint{368775} &  ludcmp     & \numprint{357320}  & \numprint{701680}\\
mvt        & \numprint{200800}  & \numprint{320000} &  gesummv    & \numprint{376000}  & \numprint{500500}\\
jacobi-1d  & \numprint{239202}  & \numprint{398000} &  syrk       & \numprint{594480}  & \numprint{975240}\\
trisolv    & \numprint{240600}  & \numprint{320000} &  adi        & \numprint{596695}  & \numprint{1059590}\\
gemm       & \numprint{1026800} & \numprint{1684200} &&\\
\hline
\end{tabular}
\vspace*{.25cm}
\caption{Basic properties of the our benchmark instances.}
        \vspace*{-.5cm}
 	\label{tab:test_instances}
\end{table}

 \section{Detailed per Instance Results}
\begin{table}
\centering
\begin{tabular}{lr|rrr||rrr||rrr}
 & & \multicolumn{3}{c}{Evolutionary Algorithm} & \multicolumn{3}{c}{Multi-Level Algorithm} & \multicolumn{3}{c}{Single-Level Algorithm} \\
graph & $k$ & Avg. Cut & Best Cut & Balance & Avg. Cut & Best Cut & Balance &Avg. Cut & Best Cut & Balance \\
\hline
\hline
2mm0                                    & 2   & \numprint{200}                           & \numprint{200}                            & \numprint{1.00}                               & \numprint{200}    & \numprint{200}    & \numprint{1.00} & \numprint{400}    & \numprint{400}    & \numprint{1.02} \\
2mm0                                    & 4   & \numprint{947}                           & \numprint{930}                            & \numprint{1.03}                               & \numprint{9167}   & \numprint{9089}   & \numprint{1.03} & \numprint{12590}  & \numprint{12533}  & \numprint{1.03} \\
2mm0                                    & 8   & \numprint{7181}                          & \numprint{6604}                           & \numprint{1.03}                               & \numprint{17445}  & \numprint{17374}  & \numprint{1.03} & \numprint{20259}  & \numprint{20231}  & \numprint{1.03} \\
2mm0                                    & 16  & \numprint{13330}                         & \numprint{13092}                          & \numprint{1.03}                               & \numprint{22196}  & \numprint{22125}  & \numprint{1.00} & \numprint{25671}  & \numprint{25591}  & \numprint{1.03} \\
2mm0                                    & 32  & \numprint{14583}                         & \numprint{14321}                          & \numprint{1.02}                               & \numprint{25178}  & \numprint{24962}  & \numprint{1.00} & \numprint{29237}  & \numprint{29209}  & \numprint{1.03} \\
\hline
3mm0                                    & 2   & \numprint{1000}                          & \numprint{1000}                           & \numprint{1.01}                               & \numprint{39069}  & \numprint{39053}  & \numprint{1.03} & \numprint{39057}  & \numprint{39055}  & \numprint{1.03} \\
3mm0                                    & 4   & \numprint{38722}                         & \numprint{37899}                          & \numprint{1.03}                               & \numprint{56109}  & \numprint{54192}  & \numprint{1.03} & \numprint{60795}  & \numprint{60007}  & \numprint{1.03} \\
3mm0                                    & 8   & \numprint{58129}                         & \numprint{49559}                          & \numprint{1.03}                               & \numprint{83225}  & \numprint{83006}  & \numprint{1.03} & \numprint{90627}  & \numprint{90449}  & \numprint{1.03} \\
3mm0                                    & 16  & \numprint{64384}                         & \numprint{60127}                          & \numprint{1.03}                               & \numprint{95052}  & \numprint{94761}  & \numprint{1.03} & \numprint{105627} & \numprint{105122} & \numprint{1.03} \\
3mm0                                    & 32  & \numprint{62279}                         & \numprint{58190}                          & \numprint{1.03}                               & \numprint{103344} & \numprint{103314} & \numprint{1.03} & \numprint{115138} & \numprint{114853} & \numprint{1.03} \\
\hline
adi                                     & 2   & \numprint{134945}                        & \numprint{134675}                         & \numprint{1.03}                               & \numprint{155232} & \numprint{155232} & \numprint{1.02} & \numprint{158115} & \numprint{158058} & \numprint{1.00} \\
adi                                     & 4   & \numprint{284666}                        & \numprint{283892}                         & \numprint{1.03}                               & \numprint{286673} & \numprint{276213} & \numprint{1.02} & \numprint{298392} & \numprint{298355} & \numprint{1.03} \\
adi                                     & 8   & \numprint{290823}                        & \numprint{290672}                         & \numprint{1.03}                               & \numprint{296728} & \numprint{296682} & \numprint{1.03} & \numprint{309067} & \numprint{308651} & \numprint{1.03} \\
adi                                     & 16  & \numprint{326963}                        & \numprint{326923}                         & \numprint{1.03}                               & \numprint{335778} & \numprint{335373} & \numprint{1.03} & \numprint{366073} & \numprint{362382} & \numprint{1.03} \\
adi                                     & 32  & \numprint{370876}                        & \numprint{370413}                         & \numprint{1.03}                               & \numprint{378883} & \numprint{378572} & \numprint{1.03} & \numprint{413394} & \numprint{413138} & \numprint{1.03} \\
\hline
atax                                    & 2   & \numprint{47826}                         & \numprint{47424}                          & \numprint{1.03}                               & \numprint{48302}  & \numprint{48302}  & \numprint{1.00} & \numprint{61425}  & \numprint{60533}  & \numprint{1.03} \\
atax                                    & 4   & \numprint{82397}                         & \numprint{76245}                          & \numprint{1.03}                               & \numprint{112616} & \numprint{111295} & \numprint{1.03} & \numprint{116183} & \numprint{115807} & \numprint{1.03} \\
atax                                    & 8   & \numprint{113410}                        & \numprint{111051}                         & \numprint{1.03}                               & \numprint{129373} & \numprint{129169} & \numprint{1.03} & \numprint{144918} & \numprint{144614} & \numprint{1.03} \\
atax                                    & 16  & \numprint{127687}                        & \numprint{125146}                         & \numprint{1.03}                               & \numprint{141709} & \numprint{141052} & \numprint{1.03} & \numprint{157799} & \numprint{157541} & \numprint{1.03} \\
atax                                    & 32  & \numprint{132092}                        & \numprint{130854}                         & \numprint{1.03}                               & \numprint{147416} & \numprint{147028} & \numprint{1.03} & \numprint{167963} & \numprint{167756} & \numprint{1.03} \\
\hline
covariance                              & 2   & \numprint{66520}                         & \numprint{66445}                          & \numprint{1.03}                               & \numprint{66432}  & \numprint{66365}  & \numprint{1.03} & \numprint{67534}  & \numprint{67450}  & \numprint{1.03} \\
covariance                              & 4   & \numprint{84626}                         & \numprint{84213}                          & \numprint{1.03}                               & \numprint{90582}  & \numprint{90170}  & \numprint{1.03} & \numprint{95801}  & \numprint{95676}  & \numprint{1.03} \\
covariance                              & 8   & \numprint{103710}                        & \numprint{102425}                         & \numprint{1.03}                               & \numprint{110996} & \numprint{109307} & \numprint{1.03} & \numprint{122410} & \numprint{122017} & \numprint{1.03} \\
covariance                              & 16  & \numprint{125816}                        & \numprint{123276}                         & \numprint{1.03}                               & \numprint{141706} & \numprint{141142} & \numprint{1.03} & \numprint{155390} & \numprint{154446} & \numprint{1.03} \\
covariance                              & 32  & \numprint{142214}                        & \numprint{137905}                         & \numprint{1.03}                               & \numprint{168378} & \numprint{167678} & \numprint{1.03} & \numprint{173512} & \numprint{173275} & \numprint{1.03} \\
\hline
doitgen                                 & 2   & \numprint{43807}                         & \numprint{42208}                          & \numprint{1.03}                               & \numprint{58218}  & \numprint{58123}  & \numprint{1.03} & \numprint{58216}  & \numprint{58190}  & \numprint{1.03} \\
doitgen                                 & 4   & \numprint{72115}                         & \numprint{71072}                          & \numprint{1.03}                               & \numprint{83422}  & \numprint{83278}  & \numprint{1.03} & \numprint{85531}  & \numprint{85279}  & \numprint{1.03} \\
doitgen                                 & 8   & \numprint{76977}                         & \numprint{75114}                          & \numprint{1.03}                               & \numprint{98418}  & \numprint{98234}  & \numprint{1.03} & \numprint{105182} & \numprint{105027} & \numprint{1.03} \\
doitgen                                 & 16  & \numprint{84203}                         & \numprint{77436}                          & \numprint{1.03}                               & \numprint{107795} & \numprint{107720} & \numprint{1.03} & \numprint{115506} & \numprint{115152} & \numprint{1.03} \\
doitgen                                 & 32  & \numprint{94135}                         & \numprint{92739}                          & \numprint{1.03}                               & \numprint{114439} & \numprint{114241} & \numprint{1.03} & \numprint{124564} & \numprint{124457} & \numprint{1.03} \\
\hline
durbin                                  & 2   & \numprint{12997}                         & \numprint{12997}                          & \numprint{1.02}                               & \numprint{13203}  & \numprint{13203}  & \numprint{1.00} & \numprint{13203}  & \numprint{13203}  & \numprint{1.00} \\
durbin                                  & 4   & \numprint{21641}                         & \numprint{21641}                          & \numprint{1.02}                               & \numprint{21724}  & \numprint{21720}  & \numprint{1.00} & \numprint{21732}  & \numprint{21730}  & \numprint{1.00} \\
durbin                                  & 8   & \numprint{27571}                         & \numprint{27571}                          & \numprint{1.01}                               & \numprint{27650}  & \numprint{27647}  & \numprint{1.01} & \numprint{27668}  & \numprint{27666}  & \numprint{1.01} \\
durbin                                  & 16  & \numprint{32865}                         & \numprint{32865}                          & \numprint{1.03}                               & \numprint{33065}  & \numprint{33045}  & \numprint{1.03} & \numprint{33340}  & \numprint{33329}  & \numprint{1.03} \\
durbin                                  & 32  & \numprint{39726}                         & \numprint{39725}                          & \numprint{1.03}                               & \numprint{40481}  & \numprint{40457}  & \numprint{1.03} & \numprint{41204}  & \numprint{41178}  & \numprint{1.03} \\
\hline
fdtd-2d                                 & 2   & \numprint{5494}                          & \numprint{5494}                           & \numprint{1.01}                               & \numprint{5966}   & \numprint{5946}   & \numprint{1.01} & \numprint{6437}   & \numprint{6427}   & \numprint{1.00} \\
fdtd-2d                                 & 4   & \numprint{15100}                         & \numprint{15099}                          & \numprint{1.03}                               & \numprint{16948}  & \numprint{16893}  & \numprint{1.02} & \numprint{18210}  & \numprint{18170}  & \numprint{1.00} \\
fdtd-2d                                 & 8   & \numprint{33087}                         & \numprint{32355}                          & \numprint{1.03}                               & \numprint{38767}  & \numprint{38687}  & \numprint{1.03} & \numprint{41267}  & \numprint{41229}  & \numprint{1.01} \\
fdtd-2d                                 & 16  & \numprint{35714}                         & \numprint{35239}                          & \numprint{1.02}                               & \numprint{78458}  & \numprint{78311}  & \numprint{1.03} & \numprint{83498}  & \numprint{83437}  & \numprint{1.03} \\
fdtd-2d                                 & 32  & \numprint{43961}                         & \numprint{42507}                          & \numprint{1.02}                               & \numprint{106003} & \numprint{105885} & \numprint{1.03} & \numprint{128443} & \numprint{128146} & \numprint{1.03} \\
\hline
gemm                                    & 2   & \numprint{383084}                        & \numprint{382433}                         & \numprint{1.03}                               & \numprint{384778} & \numprint{384490} & \numprint{1.03} & \numprint{388243} & \numprint{387685} & \numprint{1.03} \\
gemm                                    & 4   & \numprint{507250}                        & \numprint{500526}                         & \numprint{1.03}                               & \numprint{532558} & \numprint{531419} & \numprint{1.03} & \numprint{555800} & \numprint{555541} & \numprint{1.03} \\
gemm                                    & 8   & \numprint{578951}                        & \numprint{575004}                         & \numprint{1.03}                               & \numprint{611551} & \numprint{609528} & \numprint{1.03} & \numprint{649641} & \numprint{647955} & \numprint{1.03} \\
gemm                                    & 16  & \numprint{615342}                        & \numprint{613373}                         & \numprint{1.03}                               & \numprint{658565} & \numprint{654826} & \numprint{1.03} & \numprint{701624} & \numprint{699215} & \numprint{1.03} \\
gemm                                    & 32  & \numprint{626472}                        & \numprint{623271}                         & \numprint{1.03}                               & \numprint{703613} & \numprint{701886} & \numprint{1.03} & \numprint{751441} & \numprint{750144} & \numprint{1.03} \\
\hline
gemver                                  & 2   & \numprint{29349}                         & \numprint{29270}                          & \numprint{1.03}                               & \numprint{31482}  & \numprint{31430}  & \numprint{1.03} & \numprint{32785}  & \numprint{32718}  & \numprint{1.03} \\
gemver                                  & 4   & \numprint{49361}                         & \numprint{49229}                          & \numprint{1.03}                               & \numprint{54884}  & \numprint{54683}  & \numprint{1.03} & \numprint{58920}  & \numprint{58886}  & \numprint{1.03} \\
gemver                                  & 8   & \numprint{68163}                         & \numprint{67094}                          & \numprint{1.03}                               & \numprint{74114}  & \numprint{74005}  & \numprint{1.03} & \numprint{82140}  & \numprint{81935}  & \numprint{1.03} \\
gemver                                  & 16  & \numprint{78115}                         & \numprint{75596}                          & \numprint{1.03}                               & \numprint{86623}  & \numprint{86476}  & \numprint{1.02} & \numprint{98061}  & \numprint{97851}  & \numprint{1.03} \\
gemver                                  & 32  & \numprint{85331}                         & \numprint{84865}                          & \numprint{1.03}                               & \numprint{94574}  & \numprint{94295}  & \numprint{1.02} & \numprint{110439} & \numprint{110250} & \numprint{1.03} \\
\hline
gesummv                                 & 2   & \numprint{1666}                          & \numprint{500}                            & \numprint{1.02}                               & \numprint{61764}  & \numprint{61404}  & \numprint{1.03} & \numprint{102406} & \numprint{101530} & \numprint{1.01} \\
gesummv                                 & 4   & \numprint{98542}                         & \numprint{94493}                          & \numprint{1.02}                               & \numprint{109121} & \numprint{108200} & \numprint{1.00} & \numprint{135352} & \numprint{134783} & \numprint{1.03} \\
gesummv                                 & 8   & \numprint{101533}                        & \numprint{98982}                          & \numprint{1.01}                               & \numprint{116534} & \numprint{116167} & \numprint{1.01} & \numprint{159982} & \numprint{159456} & \numprint{1.03} \\
gesummv                                 & 16  & \numprint{112064}                        & \numprint{104866}                         & \numprint{1.03}                               & \numprint{123615} & \numprint{121960} & \numprint{1.02} & \numprint{184950} & \numprint{184645} & \numprint{1.03} \\
gesummv                                 & 32  & \numprint{117752}                        & \numprint{114812}                         & \numprint{1.03}                               & \numprint{135491} & \numprint{133445} & \numprint{1.03} & \numprint{195511} & \numprint{195483} & \numprint{1.03} \\
\hline
\end{tabular}
\caption{Detailed per Instance Results}
\label{tab:detailedone}
\end{table}
\begin{table}
\centering
\begin{tabular}{lr|rrr||rrr||rrr}
                                        &     & \multicolumn{3}{c}{Evolutionary Algorithm} & \multicolumn{3}{c}{Multi-Level Algorithm} & \multicolumn{3}{c}{Single-Level Algorithm} \\
graph                                   & $k$ & Avg. Cut                                   & Best Cut                                  & Balance                                       & Avg. Cut            & Best Cut          & Balance         & Avg. Cut            & Best Cut          & Balance \\
\hline
\hline
heat-3d                                 & 2   & \numprint{8695}                          & \numprint{8684}                           & \numprint{1.01}                               & \numprint{8997}   & \numprint{8975}   & \numprint{1.01} & \numprint{9136}   & \numprint{9100}   & \numprint{1.01} \\
heat-3d                                 & 4   & \numprint{14592}                         & \numprint{14592}                          & \numprint{1.01}                               & \numprint{16150}  & \numprint{16092}  & \numprint{1.02} & \numprint{16639}  & \numprint{16602}  & \numprint{1.02} \\
heat-3d                                 & 8   & \numprint{20608}                         & \numprint{20608}                          & \numprint{1.02}                               & \numprint{24869}  & \numprint{24787}  & \numprint{1.03} & \numprint{26072}  & \numprint{26024}  & \numprint{1.03} \\
heat-3d                                 & 16  & \numprint{31615}                         & \numprint{31500}                          & \numprint{1.03}                               & \numprint{41120}  & \numprint{41049}  & \numprint{1.03} & \numprint{43434}  & \numprint{43323}  & \numprint{1.03} \\
heat-3d                                 & 32  & \numprint{51963}                         & \numprint{50758}                          & \numprint{1.03}                               & \numprint{70598}  & \numprint{70524}  & \numprint{1.03} & \numprint{78086}  & \numprint{77888}  & \numprint{1.03} \\
\hline
jacobi-1d                               & 2   & \numprint{596}                           & \numprint{596}                            & \numprint{1.01}                               & \numprint{656}    & \numprint{652}    & \numprint{1.00} & \numprint{732}    & \numprint{729}    & \numprint{1.00} \\
jacobi-1d                               & 4   & \numprint{1493}                          & \numprint{1492}                           & \numprint{1.01}                               & \numprint{1739}   & \numprint{1736}   & \numprint{1.00} & \numprint{1994}   & \numprint{1990}   & \numprint{1.00} \\
jacobi-1d                               & 8   & \numprint{3136}                          & \numprint{3136}                           & \numprint{1.01}                               & \numprint{3811}   & \numprint{3803}   & \numprint{1.00} & \numprint{4398}   & \numprint{4392}   & \numprint{1.00} \\
jacobi-1d                               & 16  & \numprint{6340}                          & \numprint{6338}                           & \numprint{1.01}                               & \numprint{7884}   & \numprint{7880}   & \numprint{1.00} & \numprint{9161}   & \numprint{9159}   & \numprint{1.00} \\
jacobi-1d                               & 32  & \numprint{8923}                          & \numprint{8750}                           & \numprint{1.03}                               & \numprint{15989}  & \numprint{15987}  & \numprint{1.02} & \numprint{18613}  & \numprint{18602}  & \numprint{1.01} \\
\hline
jacobi-2d                               & 2   & \numprint{2994}                          & \numprint{2991}                           & \numprint{1.02}                               & \numprint{3227}   & \numprint{3223}   & \numprint{1.01} & \numprint{3573}   & \numprint{3568}   & \numprint{1.01} \\
jacobi-2d                               & 4   & \numprint{5701}                          & \numprint{5700}                           & \numprint{1.02}                               & \numprint{6771}   & \numprint{6749}   & \numprint{1.01} & \numprint{7808}   & \numprint{7797}   & \numprint{1.01} \\
jacobi-2d                               & 8   & \numprint{9417}                          & \numprint{9416}                           & \numprint{1.03}                               & \numprint{12287}  & \numprint{12160}  & \numprint{1.03} & \numprint{14714}  & \numprint{14699}  & \numprint{1.03} \\
jacobi-2d                               & 16  & \numprint{16274}                         & \numprint{16231}                          & \numprint{1.03}                               & \numprint{23070}  & \numprint{22971}  & \numprint{1.03} & \numprint{27943}  & \numprint{27860}  & \numprint{1.03} \\
jacobi-2d                               & 32  & \numprint{22181}                         & \numprint{21758}                          & \numprint{1.03}                               & \numprint{43956}  & \numprint{43928}  & \numprint{1.03} & \numprint{52988}  & \numprint{52969}  & \numprint{1.03} \\
\hline
lu                                      & 2   & \numprint{5210}                          & \numprint{5162}                           & \numprint{1.03}                               & \numprint{5183}   & \numprint{5174}   & \numprint{1.03} & \numprint{5184}   & \numprint{5173}   & \numprint{1.03} \\
lu                                      & 4   & \numprint{13528}                         & \numprint{13510}                          & \numprint{1.03}                               & \numprint{14160}  & \numprint{14122}  & \numprint{1.03} & \numprint{14220}  & \numprint{14189}  & \numprint{1.03} \\
lu                                      & 8   & \numprint{33307}                         & \numprint{33211}                          & \numprint{1.03}                               & \numprint{33890}  & \numprint{33764}  & \numprint{1.03} & \numprint{33722}  & \numprint{33625}  & \numprint{1.03} \\
lu                                      & 16  & \numprint{74543}                         & \numprint{74006}                          & \numprint{1.03}                               & \numprint{76399}  & \numprint{75043}  & \numprint{1.03} & \numprint{78698}  & \numprint{78165}  & \numprint{1.03} \\
lu                                      & 32  & \numprint{130674}                        & \numprint{129954}                         & \numprint{1.03}                               & \numprint{143735} & \numprint{143396} & \numprint{1.03} & \numprint{151452} & \numprint{150549} & \numprint{1.02} \\
\hline
ludcmp                                  & 2   & \numprint{5380}                          & \numprint{5337}                           & \numprint{1.02}                               & \numprint{5337}   & \numprint{5337}   & \numprint{1.02} & \numprint{5337}   & \numprint{5337}   & \numprint{1.02} \\
ludcmp                                  & 4   & \numprint{14744}                         & \numprint{14744}                          & \numprint{1.03}                               & \numprint{17352}  & \numprint{17278}  & \numprint{1.03} & \numprint{17322}  & \numprint{17252}  & \numprint{1.03} \\
ludcmp                                  & 8   & \numprint{37228}                         & \numprint{37069}                          & \numprint{1.03}                               & \numprint{40579}  & \numprint{40420}  & \numprint{1.03} & \numprint{40164}  & \numprint{39752}  & \numprint{1.03} \\
ludcmp                                  & 16  & \numprint{78646}                         & \numprint{78467}                          & \numprint{1.03}                               & \numprint{81951}  & \numprint{81778}  & \numprint{1.03} & \numprint{85882}  & \numprint{85582}  & \numprint{1.03} \\
ludcmp                                  & 32  & \numprint{134758}                        & \numprint{134288}                         & \numprint{1.03}                               & \numprint{150112} & \numprint{149930} & \numprint{1.03} & \numprint{157788} & \numprint{156570} & \numprint{1.03} \\
\hline
mvt                                     & 2   & \numprint{24528}                         & \numprint{23091}                          & \numprint{1.02}                               & \numprint{63485}  & \numprint{63054}  & \numprint{1.03} & \numprint{80468}  & \numprint{80408}  & \numprint{1.03} \\
mvt                                     & 4   & \numprint{74386}                         & \numprint{73035}                          & \numprint{1.02}                               & \numprint{83951}  & \numprint{82868}  & \numprint{1.03} & \numprint{102122} & \numprint{101359} & \numprint{1.03} \\
mvt                                     & 8   & \numprint{86525}                         & \numprint{82221}                          & \numprint{1.03}                               & \numprint{96695}  & \numprint{96362}  & \numprint{1.01} & \numprint{116068} & \numprint{115722} & \numprint{1.03} \\
mvt                                     & 16  & \numprint{99144}                         & \numprint{97941}                          & \numprint{1.03}                               & \numprint{107347} & \numprint{107032} & \numprint{1.01} & \numprint{129178} & \numprint{128962} & \numprint{1.03} \\
mvt                                     & 32  & \numprint{105066}                        & \numprint{104917}                         & \numprint{1.03}                               & \numprint{115123} & \numprint{114845} & \numprint{1.01} & \numprint{143436} & \numprint{143205} & \numprint{1.03} \\
\hline
seidel-2d                               & 2   & \numprint{4991}                          & \numprint{4969}                           & \numprint{1.01}                               & \numprint{5441}   & \numprint{5384}   & \numprint{1.00} & \numprint{5461}   & \numprint{5397}   & \numprint{1.00} \\
seidel-2d                               & 4   & \numprint{12197}                         & \numprint{12169}                          & \numprint{1.01}                               & \numprint{13358}  & \numprint{13334}  & \numprint{1.01} & \numprint{13387}  & \numprint{13372}  & \numprint{1.01} \\
seidel-2d                               & 8   & \numprint{21419}                         & \numprint{21400}                          & \numprint{1.01}                               & \numprint{24011}  & \numprint{23958}  & \numprint{1.02} & \numprint{24167}  & \numprint{24150}  & \numprint{1.01} \\
seidel-2d                               & 16  & \numprint{38222}                         & \numprint{38110}                          & \numprint{1.02}                               & \numprint{43169}  & \numprint{43071}  & \numprint{1.02} & \numprint{43500}  & \numprint{43419}  & \numprint{1.02} \\
seidel-2d                               & 32  & \numprint{52246}                         & \numprint{51531}                          & \numprint{1.03}                               & \numprint{79882}  & \numprint{79813}  & \numprint{1.03} & \numprint{81433}  & \numprint{81077}  & \numprint{1.03} \\
\hline
symm                                    & 2   & \numprint{94357}                         & \numprint{94214}                          & \numprint{1.03}                               & \numprint{95630}  & \numprint{95429}  & \numprint{1.03} & \numprint{96934}  & \numprint{96765}  & \numprint{1.03} \\
symm                                    & 4   & \numprint{127497}                        & \numprint{126207}                         & \numprint{1.03}                               & \numprint{134923} & \numprint{134888} & \numprint{1.03} & \numprint{149064} & \numprint{148653} & \numprint{1.03} \\
symm                                    & 8   & \numprint{152984}                        & \numprint{151168}                         & \numprint{1.03}                               & \numprint{161622} & \numprint{161575} & \numprint{1.03} & \numprint{175299} & \numprint{175169} & \numprint{1.03} \\
symm                                    & 16  & \numprint{167822}                        & \numprint{167512}                         & \numprint{1.03}                               & \numprint{177001} & \numprint{176568} & \numprint{1.03} & \numprint{190628} & \numprint{190519} & \numprint{1.03} \\
symm                                    & 32  & \numprint{174938}                        & \numprint{174843}                         & \numprint{1.03}                               & \numprint{185321} & \numprint{185207} & \numprint{1.03} & \numprint{207974} & \numprint{207694} & \numprint{1.03} \\
\hline
syr2k                                   & 2   & \numprint{11098}                         & \numprint{3894}                           & \numprint{1.03}                               & \numprint{35756}  & \numprint{35731}  & \numprint{1.03} & \numprint{36841}  & \numprint{36708}  & \numprint{1.03} \\
syr2k                                   & 4   & \numprint{49662}                         & \numprint{48021}                          & \numprint{1.03}                               & \numprint{52430}  & \numprint{52388}  & \numprint{1.03} & \numprint{56695}  & \numprint{56589}  & \numprint{1.03} \\
syr2k                                   & 8   & \numprint{57584}                         & \numprint{57408}                          & \numprint{1.03}                               & \numprint{60321}  & \numprint{60237}  & \numprint{1.03} & \numprint{64928}  & \numprint{64825}  & \numprint{1.03} \\
syr2k                                   & 16  & \numprint{59780}                         & \numprint{59594}                          & \numprint{1.03}                               & \numprint{64880}  & \numprint{64791}  & \numprint{1.03} & \numprint{70880}  & \numprint{70792}  & \numprint{1.03} \\
syr2k                                   & 32  & \numprint{60502}                         & \numprint{60085}                          & \numprint{1.03}                               & \numprint{67932}  & \numprint{67900}  & \numprint{1.03} & \numprint{77239}  & \numprint{77206}  & \numprint{1.03} \\
\hline
syrk                                    & 2   & \numprint{219263}                        & \numprint{218019}                         & \numprint{1.03}                               & \numprint{220692} & \numprint{220530} & \numprint{1.03} & \numprint{222919} & \numprint{222696} & \numprint{1.03} \\
syrk                                    & 4   & \numprint{289509}                        & \numprint{289088}                         & \numprint{1.03}                               & \numprint{300418} & \numprint{299777} & \numprint{1.03} & \numprint{317979} & \numprint{317756} & \numprint{1.03} \\
syrk                                    & 8   & \numprint{329466}                        & \numprint{327712}                         & \numprint{1.03}                               & \numprint{341826} & \numprint{341368} & \numprint{1.03} & \numprint{371901} & \numprint{369820} & \numprint{1.03} \\
syrk                                    & 16  & \numprint{354223}                        & \numprint{351824}                         & \numprint{1.03}                               & \numprint{366694} & \numprint{366500} & \numprint{1.03} & \numprint{402556} & \numprint{401806} & \numprint{1.03} \\
syrk                                    & 32  & \numprint{362016}                        & \numprint{359544}                         & \numprint{1.03}                               & \numprint{396365} & \numprint{394132} & \numprint{1.03} & \numprint{431733} & \numprint{431250} & \numprint{1.03} \\
\hline
trisolv                                 & 2   & \numprint{6788}                          & \numprint{3549}                           & \numprint{1.03}                               & \numprint{27767}  & \numprint{27181}  & \numprint{1.03} & \numprint{46291}  & \numprint{46257}  & \numprint{1.03} \\
trisolv                                 & 4   & \numprint{43927}                         & \numprint{43549}                          & \numprint{1.03}                               & \numprint{45436}  & \numprint{44649}  & \numprint{1.03} & \numprint{55527}  & \numprint{55476}  & \numprint{1.03} \\
trisolv                                 & 8   & \numprint{66148}                         & \numprint{65662}                          & \numprint{1.03}                               & \numprint{66187}  & \numprint{65420}  & \numprint{1.03} & \numprint{68497}  & \numprint{68395}  & \numprint{1.03} \\
trisolv                                 & 16  & \numprint{71838}                         & \numprint{71447}                          & \numprint{1.03}                               & \numprint{72206}  & \numprint{72202}  & \numprint{1.03} & \numprint{72966}  & \numprint{72957}  & \numprint{1.03} \\
trisolv                                 & 32  & \numprint{79125}                         & \numprint{79071}                          & \numprint{1.03}                               & \numprint{79173}  & \numprint{79103}  & \numprint{1.03} & \numprint{79793}  & \numprint{79679}  & \numprint{1.03} \\
\hline
trmm                                    & 2   & \numprint{138937}                        & \numprint{138725}                         & \numprint{1.03}                               & \numprint{139245} & \numprint{139188} & \numprint{1.03} & \numprint{139273} & \numprint{139259} & \numprint{1.03} \\
trmm                                    & 4   & \numprint{192752}                        & \numprint{191492}                         & \numprint{1.03}                               & \numprint{200570} & \numprint{200232} & \numprint{1.03} & \numprint{208334} & \numprint{208057} & \numprint{1.03} \\
trmm                                    & 8   & \numprint{225192}                        & \numprint{223529}                         & \numprint{1.03}                               & \numprint{238791} & \numprint{238337} & \numprint{1.03} & \numprint{260719} & \numprint{259607} & \numprint{1.03} \\
trmm                                    & 16  & \numprint{240788}                        & \numprint{238159}                         & \numprint{1.03}                               & \numprint{261560} & \numprint{261173} & \numprint{1.03} & \numprint{287082} & \numprint{286768} & \numprint{1.03} \\
trmm                                    & 32  & \numprint{246407}                        & \numprint{245173}                         & \numprint{1.03}                               & \numprint{281417} & \numprint{281242} & \numprint{1.03} & \numprint{300631} & \numprint{299939} & \numprint{1.03} \\
\hline
\end{tabular}
\caption{Detailed per Instance Results}
\label{tab:detailedtwo}
\end{table}
\end{appendix}
\end{document}